\documentclass[review=false]{jfp-epi}
\usepackage{jfp-cup2epi}
\usepackage[utf8]{inputenc}
\usepackage{amsthm}
\usepackage{thmtools} %declaring new theorems, listing all theorems etc. Needs amsthm to work. http://ftp.gwdg.de/pub/ctan/macros/latex/exptl/thmtools/thmtools.pdf
\usepackage{thm-restate} %Allows to restate a theorem later in the document part of thmtools but needs to be imported separately http://ftp.gwdg.de/pub/ctan/macros/latex/exptl/thmtools/thmtools.pdf
\usepackage{xifthen} % allows additional tests in ifthenelse http://ctan.space-pro.be/tex-archive/macros/latex/contrib/xifthen/xifthen.pdf

\usepackage{xcolor} %various colour commands http://ftp.fernuni-hagen.de/ftp-dir/pub/mirrors/www.ctan.org/macros/latex/contrib/xcolor/xcolor.pdf
\usepackage{soul} % allows colored strikethoughs
\setstcolor{red}
\usepackage{xparse}

\usepackage[english=british]{csquotes} %Allows for easier usage various kinds of quotes. http://ftp.fau.de/ctan/macros/latex/contrib/csquotes/csquotes.pdf

\theoremstyle{definition}
\declaretheorem[name=Theorem,numberwithin=section]{theorem}
\declaretheorem[name=Lemma,numberwithin=section]{lemma}

\declaretheorem[name=Definition,numberwithin=section]{definition}
\declaretheorem[name=Definition,unnumbered]{definition*}

\usepackage{mathpartir} % For typesetting inference rules http://ctan.mirror.norbert-ruehl.de/macros/latex/contrib/mathpartir/mathpartir.pdf
\usepackage{xargs} % Allows do define commands with more than one optional argument https://ftp.agdsn.de/pub/mirrors/latex/dante/macros/latex/contrib/xargs/xargs.pdf
\usepackage{mathtools} % Extension to amsmath, if used amsmath does not need to be loaded again. http://ctan.mirror.norbert-ruehl.de/macros/latex/contrib/mathtools/mathtools.pdf
\usepackage{marvosym} % Symbols fond Does for example seem to provide the \Lightning symbol https://www.ctan.org/tex-archive/fonts/marvosym
\usepackage{eulervm
} % Maths symbols, makes font in math mode much more readable

\DeclareMathSymbol{\alpha}  {\mathord}{letters}{"0B}
\DeclareMathSymbol{\beta}   {\mathord}{letters}{"0C}
\DeclareMathSymbol{\gamma}  {\mathord}{letters}{"0D}
\DeclareMathSymbol{\delta}  {\mathord}{letters}{"0E}
\DeclareMathSymbol{\epsilon}{\mathord}{letters}{"0F}
\DeclareMathSymbol{\zeta}   {\mathord}{letters}{"10}
\DeclareMathSymbol{\eta}    {\mathord}{letters}{"11}
\DeclareMathSymbol{\theta}  {\mathord}{letters}{"12}
\DeclareMathSymbol{\iota}   {\mathord}{letters}{"13}
\DeclareMathSymbol{\kappa}  {\mathord}{letters}{"14}
\DeclareMathSymbol{\lambda} {\mathord}{letters}{"15}
\DeclareMathSymbol{\mu}     {\mathord}{letters}{"16}
\DeclareMathSymbol{\nu}     {\mathord}{letters}{"17}
\DeclareMathSymbol{\xi}     {\mathord}{letters}{"18}
\DeclareMathSymbol{\pi}     {\mathord}{letters}{"19}
\DeclareMathSymbol{\rho}    {\mathord}{letters}{"1A}
\DeclareMathSymbol{\sigma}  {\mathord}{letters}{"1B}
\DeclareMathSymbol{\tau}    {\mathord}{letters}{"1C}
\DeclareMathSymbol{\upsilon}{\mathord}{letters}{"1D}
\DeclareMathSymbol{\phi}    {\mathord}{letters}{"1E}
\DeclareMathSymbol{\chi}    {\mathord}{letters}{"1F}
\DeclareMathSymbol{\psi}    {\mathord}{letters}{"20}
\DeclareMathSymbol{\omega}  {\mathord}{letters}{"21}
\DeclareMathSymbol{\varepsilon}{\mathord}{letters}{"22}
\DeclareMathSymbol{\vartheta}{\mathord}{letters}{"23}
\DeclareMathSymbol{\varpi}  {\mathord}{letters}{"24}
\DeclareMathSymbol{\varphi} {\mathord}{letters}{"27}

\newcommand{\curly}[1]{\left\lbrace#1\right\rbrace}

\newcommand{\set}[1]{\curly{#1}}
\newcommand{\where}{\,\middle|\,}
\newcommand{\NN}{\mathbb{N}}

\newcommand{\hyp}[2]{{#1}:{#2}}

\newcommand{\Infer}[3]{\inferrule*[right={\textbf{#1}}]{#2}{#3}}
\newcommand{\axiom}[2]{\inferrule*[right={\textbf{#1}}]{ }{#2}}

\newcommand{\wf}[1][\Gamma]{\triangleright_{#1}}
\newcommand{\bwf}{\overset{\AttackerA}{\triangleright}}

\newcommand{\unarybase}[1]{\lceil #1 \rceil}
\newcommand{\vs}[1]{\unarybase{#1} _{\mathcal{V}}}
\newcommand{\es}[2]{\unarybase{#1}^{#2} _{E}}

\newcommand{\binarybase}[1]{\lceil\!\!\lceil #1 \rceil\!\!\rceil}

\newcommand{\vb}[1]{\binarybase{#1}^{\AttackerA} _{\mathcal{V}}}
\newcommand{\eb}[2][pc]{\binarybase{#2}^{\AttackerA%,#1
	}_E}

\newcommand{\setof}[1]{\left\{#1\right\}}
\newcommand{\comprehend}[2]{\setof{{#1} \;\middle|\; {#2}}}

\newcommand{\equivWorld}[1][(W,m)]{\approx^{\AttackerA}_{#1}\;}

\newcommand{\equivWelt}[1][(W,m)]{\approxeq^{\AttackerA}_{#1}}

\newcommand{\inv}[1]{inv_\AttackerA(#1)}

\newcommand{\nored}[3]{\axiom{no-red}{\Steps{#1,#2,#3,\EmptyTrace,\AttackerA, #1,#2,#3}}}

\makeatletter    
\DeclareRobustCommand\DetectUnderscore[1]{%
	\begingroup
	\protected@edef\@tempa{#1}%
	\@onelevel@sanitize\@tempa
	\expandafter\expandafter\expandafter\endgroup
	\expandafter\expandafter\expandafter\ifthenelse
	\expandafter\expandafter\expandafter{%
		\expandafter\expandafter\expandafter\isin
		\expandafter\expandafter\expandafter{%
			\expandafter\expandafter\string_%
			\expandafter}%
		\expandafter{%
			\@tempa}}{{(#1)}}{{#1}}%
}%
\newcommand{\low}[1]{%
	\DetectUnderscore{#1}%
	_{{\AttackerA}}%
}
\makeatother

\newcommand{\lowequiv}{\approx_\AttackerA}

\newcommand{\subst}[3][x]{#3[#1 \mapsto #2]}
\newboolean{fullversion}
\newboolean{ready}
\newcommand{\onlyfull}[1]{\ifthenelse{\boolean{fullversion}}{#1}{}}
\newcommand{\notready}[1]{\ifthenelse{\boolean{ready}}{}{#1}}

\newcommand{\btext}[1]{%
	\begingroup
	\sethlcolor{cyan}%
	\hl{#1}%
	\endgroup
}
\DeclareRobustCommand{\added}[1]{\ifthenelse{\boolean{ready}}{#1}{\btext{#1}}}
\newcommand{\addedmaths}[1]{\ifthenelse{\boolean{ready}}{\ensuremath{#1}}{\colorbox{cyan}{\ensuremath{#1}}}}

\colorlet{xtralightgray}{lightgray!65!white}
\colorlet{xtralightyellow}{yellow!40!white}
\newcommand{\newstuff}[1]{\colorbox{xtralightyellow}{\ensuremath{#1}}}
\newcommand{\unimp}[1]{\colorbox{xtralightgray}{\ensuremath{#1}}}
\newcommand{\newtext}[1]{\smash{\colorbox{xtralightyellow}{\hspace{-0.25em}#1\hspace{-0.25em}}}}
\newcommand{\unimptext}[1]{\smash{\colorbox{xtralightgray}{\hspace{-0.25em}#1\hspace{-0.25em}}}}
\DeclareRobustCommand{\del}[1]{\notready{\st{#1}}}
\DeclareRobustCommand{\delmaths}[1]{\notready{\colorbox{red}{\ensuremath{#1}}}}
\newcommand{\mnb}[1]{\medskip\noindent\textbf{#1.}}
\newcommand{\calcname}{\mbox{$\lambda$-\small{WHR}}}
\NewDocumentCommand{\rname}{m}{\textbf{\textsc{#1}}}

\newcommand{\ebb}[2][pc]{\binarybase{#2}^{\AttackerA%,#1
	}_{E_{\delmaths{\beta}\addedmaths{r}}}}
\newcommand{\esb}[2]{\unarybase{#1}^{#2} _{E_{\delmaths{\beta}\addedmaths{r}}}}

\usepackage{FG}
\usepackage{pl-syntax}
\setboolean{fullversion}{false}
\setboolean{ready}{true}
\usepackage[T1]{fontenc}
\usepackage{subfig}

\begin{document}
\journaltitle{JFP}
\cpr{Cambridge University Press}
\doival{10.1017/xxxxx}

%\lefttitle{\LaTeX\ Supplement}
%\righttitle{Journal of Functional Programming}

\totalpg{\pageref{lastpage01}}
\jnlDoiYr{2025}

\title%[Short Title]
{Compositional security definitions for higher-order where declassification}    

%\begin{authgrp}
\author{Jan Menz}
%\authornote{with author1 note}          %% \authornote is optional;                                       %% can be repeated if necessary
\orcid{0009-0004-6821-1987}             %% \orcid is optional
\affiliation{
\institution{Max Planck Institute for Software Systems \\
Saarland Informatics Campus}
\city{66123 Saarbrücken}
\country{Germany}
\authoremail{jmenz@mpi-sws.org}
}
%% Author with two affiliations and emails.
\author{Andrew K. Hirsch}
%\authornote{with author2 note}          %% \authornote is optional;
%                                     %% can be repeated if necessary
\orcid{0000-0003-2518-614X}             %% \orcid is optional
\affiliation{
\institution{Department of Computer Science and Engineering   \\
University at Buffalo, SUNY\\
113M Davis Hall}\\
\city{14260-2500 Buffalo New York}
\country{USA}                  % \country is recommended
\authoremail{akhirsch@buffalo.edu}         %% \email is recommended
}
\author{Peixuan Li}
%\authornote{}          %% \authornote is optional;
%% can be repeated if necessary
\orcid{0009-0005-9392-3481}             %% \orcid is optional
\affiliation{
\institution{Pennsylvania State University}           %% \institution is required
\city{16802 University Park Pennsylvania}
\country{USA}
\authoremail{pzl129@cse.psu.edu}          %% \email is recommended
}
\author{Deepak Garg}
%\authornote{with author1 note}          %% \authornote is optional;
%% can be repeated if necessary
\orcid{0000-0002-0888-3093}             %% \orcid is optional
\affiliation{
	\institution{Max Planck Institute for Software Systems \\
	Saarland Informatics Campus}
	\city{66123 Saarbrücken}
	\country{Germany}
	\authoremail{dg@mpi-sws.org}          %% \email is recommended
}
%\end{authgrp}
% Abstract
% Note: \begin{abstract}...\end{abstract} environment must come
% before \maketitle command
\begin{abstract}
To ensure programs do not leak private data, we often want to be able to
provide formal guarantees ensuring such data is handled correctly.
Often, we cannot keep such data secret entirely; instead programmers
specify how private data may be \emph{declassified}.  While security
definitions for declassification exist, they mostly do not handle
higher-order programs.  In fact, in the higher-order setting no
compositional security definition exists for intensional
information-flow properties such as \emph{where} declassification,
which allows declassification in specific parts of a program.  We use
logical relations to build a model (and thus security definition) of
where declassification.  The key insight required for our model is
that we must stop enforcing indistinguishability once a \emph{relevant
declassification} has occurred.  We show that the resulting security
definition provides more security than the most related previous
definition, which is for the lower-order setting.
\added{This paper is an extended version of the paper of the same name published at OOPSLA 2023 (\mbox{\cite{MenzHLG23}}).}
\end{abstract}

\maketitle

% !TeX spellcheck = en_US
\section{Introduction}
\label{sec:introduction}

Language-based information-flow control guarantees that secret information does not flow to attacker-visible outputs of a program if such a flow would violate the security policy.
The strictest form of such a guarantee is often formalized as \emph{noninterference}, which says that a program is secure with respect to an attacker when any two runs of the program that differ only in secrets result in the same attacker-visible outcomes.

In practice, we often want less restrictive security policies since many programs intentionally reveal some secret information.
In other words, secret information may not remain secret in its entirety forever; parts of it may need to be \emph{declassified} at some points.
Hence, there is need for program-security definitions that admit declassification of information in conformance with a given policy.
While many such definitions exist for first-order programs---where functions cannot be passed to other functions and cannot be stored in memory~(\cite{BrobergS09,BrobergS10,SabelfeldM03,KozyriS20,AskarovS07A,AskarovS07B,BanerjeeNR08,MantelS04})---there are fewer such definitions for higher-order programs.

Most definitions for higher-order program security with declassification target
\emph{what~declassification}~(\cite{PengZ05, CruzRST17,NgoNR20}), wherein policies specify 
precisely what information may be revealed.
For example, a policy might specify that a secret number may not be revealed in its entirety, but its parity---whether the number is even or odd---may be revealed.
However, what~declassification is purely extensional, i.e.\added{,} based entirely on the input-output behavior, and therefore cannot specify temporal aspects -- \emph{where} and \emph{when} in a program information can be declassified.
In practice, \del{this}\added{such} aspect\added{s are} \del{is }also important.

For where~and~when~declassification, all security definitions supporting higher-order programs of which we are aware~(\cite{BrobergS06,MatosB05}) target the so-called \emph{store-based setting}.
In this setting, security labels, which represent confidentiality policies and tell us how secret the data we are dealing with is, are attached to memory locations only.
Unfortunately, definitions in the store-based setting give no meaningful notion of value security, even for those values---like function values---that have suspended computation in them.
Instead, all values are usually assumed secure by virtue of not doing any computation and security is defined only for \emph{whole-program}~expressions.

As a result, existing definitions of higher-order program security with where/when~declassification are not compositional.
By \emph{compositional} we mean ``homomorphic with respect to the programming language's constructs'', i.e., a program composed of secure partial programs should also be secure.
For example, applying a secure function to a secure argument always result in a secure expression.
However, the existing definitions in the store-based setting described above are not compositional in this sense.
To see this, consider two locations~$l$ and~$h$ labeled public and secret, respectively, along with a function $f \triangleq \Lam{x}{(\Assgn{l}{\Deref{h}})}$, which copies $h$ to $l$ ignoring its argument~$x$.
Clearly, $f$'s body is insecure because $f$ leaks $h$ without declassifying it.
However, store-based definitions deem all values secure and thus deem $f$ secure.
Next, consider the (trivially secure) argument~$5$.
If the security definition were compositional, the expression $\App{f}{5}$ should also be secure.
But, the expression $\App{f}{5}$ leaks $h$ to $l$, which is recognized as a security breach by the aforementioned store-based~definitions!

As an immediate consequence of this lack of compositionality, programs cannot be verified secure against these definitions \emph{modularly}.
Rather, one must verify whole-program security.
This constraint can be problematic.
For example, libraries may be used in many different programs which do not yet exist.
Hence, the authors of a library \emph{cannot} verify the security of their library.
Instead, each client must verify their program, including any library code they use, themselves.
This \added{duplicates a lot of work and }may \added{even }make some verification tasks---such as those for sufficiently large programs---infeasible.
Using a compositional security definition instead, we can verify each library function \emph{once}, reusing the security proof every time the library gets used.

Since type systems are usually compositional by construction, many papers side step this problem by using a type system to enforce a noncompositional security property.
This allows programmers to reuse the proof that a well typed library function is secure, just as we advocated above.
However, such type systems are necessarily conservative; they fail to recognize some secure programs as secure.
For instance, imagine that we wish to verify the security of a software library used to build online shops.
This library provides functions to calculate shipping costs.
One function sets a new shipping cost and logs the fact that the shipping cost has been set \del{(this log is for debugging)}\added{for the purpose of debugging}:
\[\WordVar{SetShippingCosts} \triangleq \Lam{\WordVar{cost}}{\Assgn{\WordVar{shipping\_cost}}{\WordVar{cost}}}; \Assgn{\WordVar{costs\_set}}{\Const{\WordVar{true}}}\]
Another function calculates and sets the shipping costs based on whether the customer is living abroad or not:
\[
\WordVar{CalculatePostage} \triangleq \Lam*{\WordVar{isAbroad}}{\Lam*{\WordVar{fee\_local}}{\Lam*{\WordVar{fee\_abroad}}{\LamNL*{\WordVar{customer}}{\ExprIf*{\App{\WordVar{isAbroad}}{\WordVar{customer}} }{\App{\WordVar{SetShippingCosts}}{\WordVar{fee\_abroad}} }{\App{\WordVar{SetShippingCosts}}{\WordVar{fee\_local}} } } } } }
\]
This function takes several parameters chosen by the client of the library and a customer of the online shop as inputs.
Because the location of the customer is sensitive data, the policies of the result of the function parameter $\WordVar{isAbroad}$ and of the variable $\WordVar{shipping\_cost}$ are more secret than the log variable $\WordVar{costs\_set}$.
While no information about $(\App{\WordVar{isAbroad}}{\WordVar{customer}})$ and $\WordVar{shipping\_cost}$ leaks to $\WordVar{costs\_set}$ in the computation of $\WordVar{CalculatePostage}$, the  $\WordVar{CalculatePostage}$~function would be rejected as insecure by most security type systems.
After all, it contains writes of less-secret data (the setting of $\WordVar{costs\_set})$ controlled by more-secret data (the result of $\App{\WordVar{isAbroad}}{\WordVar{customer}}$).
The type systems cannot determine that the location of the customer is not leaked into $\WordVar{costs\_set}$ via the writes in the two branches, so they reject the program, even though it is intuitively~secure.

Because of this shortcoming, we cannot exclusively rely on type systems for compositional verification of our programs.
Instead, we need more expressive security definitions that are compositional in their own right.
Such definitions also have an additional benefit: they allow us to apply the powerful method of \emph{semantic typing}.
In this technique, a verification engineer proves the bulk of the program secure using the type system.
Only small parts of the program that require security reasoning beyond the scope of the type system, such as the $\WordVar{CalculatePostage}$ function above, are proved secure directly against the security definition in the semantics.
The proofs done in the type system and those done directly in the semantics are then composed using the compositionality of the security definition.

\mnb{Our contribution} 
Our goal in this paper is to develop a security definition for \emph{where declassification} that is \emph{compositional} and \emph{supports higher-order programs}.
Since the store-based setting used by current definitions of where declassification leads to noncompositional definitions, we turn to an alternative setting, the \emph{value-based setting}.
In the value-based setting, values (and types) have associated security labels.
We particularly work with a \emph{fine-grained} value-based setting, where \emph{every} value and type is associated with a security~label.

Changing to the value-based setting allows us to define security via \emph{logical relations}, creating a compositional-by-construction security definition.
Existing logical-relations definitions of information-flow security define \emph{extensional} security properties---i.e., properties where security only depends on the input-output behavior of the program~\cite[see
e.g.,][]{RajaniG18,FruminKB21,GregersenBTB21,CruzRST17,NgoNR20}.
For such properties, logical relations achieve compositionality by studying how the computation suspended in higher-order values can execute---the model simply evaluates those computations completely!
For instance, given two functions, a logical-relations model of noninterference would say that those functions are indistinguishable if, given indistinguishable inputs, they give indistinguishable outputs.
However, where~and~when~declassification allow and disallow declassification based on internal state, such as the current state of the local memory ("when") or the subexpression being evaluated ("where").
These aspects cannot be observed from the outside by just analyzing the input/output behavior of the program, making them \emph{intensional} security definitions.
Consequently, depending on how the internal state changes during the execution, we can no longer expect two functions given indistinguishable inputs to produce indistinguishable results.
Hence, \textbf{a security definition for where declassification must stop enforcing indistinguishability under some circumstances.}

To model this requirement, we identify and formalize what we call \emph{relevant declassification steps}.
Relevant declassification steps are those for which we both must stop enforcing indistinguishability \emph{and} can do so securely.
This work uses relevant declassification to build a logical relation.
This insight---that we can stop enforcing indistinguishability after a relevant declassification---can, we believe, generalize to other notions of security for systems with intensional declassification.
However, since we work in a setting with higher-order state and nondeterministic allocation of memory, this formalization is quite technical.
Our main technical contribution is a description of such a formalization of relevant declassification, a logical relation built around this formalization, and a compositional security definition derived from the logical relation.

Concretely, our contributions are as follows:
first, we present {\calcname}, a higher-order programming language and security type system allowing explicit where declassification.
{\calcname} is based on a language and type system for noninterference called FG~(\cite{RajaniG18}) and extends it with declassification policies in the style of Flow Locks~(\cite{BrobergS06,BrobergS09}).
In Flow Locks, distinguished variables called \emph{locks} specify \emph{when} (i.e., in which states) a value may be declassified to an actor.
For example, the policy $\Policy{\Clause{\{\Allow\}}{\Alice}}$ says ``the actor $\Alice$ can see whatever this policy protects if the lock $\Allow$ is open.''
Policies are then used as security labels: the visibility of a value labeled with a policy is subject to that policy.
This allows value protection to depend on locks, which can be opened and closed.
Unlike Flow~Locks, the lock-opening and -closing constructs of {\calcname} are syntactically scoped for technical convenience.
Hence, whether a lock is open or not is clearly determined by the program
subexpression being executed\added{,} which means that {\calcname} policies model where declassification, rather than when declassification.

Second, we extend FG's compositional security definition to support where declassification.
FG achieves compositionality by relying on a \emph{binary logical relation}---a relational interpretation of labeled types defining when two values or expressions of a given type are indistinguishable to an adversary.
The logical relation is indexed with a (labeled) type and a Kripke world~(\cite{AhmedDR09}), which specifies types of locations in the heap.
This relation treats functions as indistinguishable exactly when they are extensionally indistinguishable, that is they produce indistinguishable results given indistinguishable inputs, forcing compositionality.
A program is secure if it is indistinguishable from itself when run with different secrets.
To handle declassification, we additionally index the relation with the open locks in the two programs and add a formalization of relevant declassification steps, which captures when we must stop enforcing indistinguishability.

Third, we prove {\calcname}'s type system sound relative to this semantic model.
This proof shows that every typing rule is semantically admissible.
That is, our typing rules remain valid if we replace syntactic typing in premises and the conclusion with semantic typing based on the logical relation.
This establishes that typing a program is a sound technique for proving it secure, and that mixing type-based and semantic proofs is sound.
It also witnesses the compositionality of the security~definition.

Finally, we connect our security definition to a prior security definition for Flow Locks~(\cite{BrobergS09}).
This security definition is based on a different kind of security model: \emph{attacker knowledge}~(\cite{AskarovS07B}).
Briefly, this security model first defines what an attacker knows based on observations it has made during the execution of a program.
Then, it deems a program secure if the attacker's knowledge does not increase except at relevant declassification events (i.e., declassification events that reveal new data to the attacker).
Flow Locks' store-based security definition is not compositional and is limited to first-order state (where functions cannot be stored in the heap), while our definition is compositional and supports higher-order state.
We prove that our security definition, restricted to first-order state, is stronger---i.e., less permissive---than the prior definition showing that compositionality does not weaken security.

For reasons of \del{space and }readability, proofs and some technical details are omitted from this paper.
The accompanying technical appendix\added{\footnote{\href{https://gitlab.mpi-sws.org/Quarkbeast/lambda-where-fullproofs/-/raw/main/Technical_Appendix.pdf}{\added{https://gitlab.mpi-sws.org/Quarkbeast/lambda-where-fullproofs/-/raw/main/Technical\_Appendix.pdf}}}}~(\cite{MenzHLG25}) contains these.

% !TeX spellcheck = en_US
\section{Language}
\label{sec:language}

\begin{figure*}
  { \small
    \begin{syntax}
      \abstractCategory[Actors]{\A}
      \abstractCategory[Locks]{\LockSigma}
      \categoryFromSet[Lock Sets]{\LockSetSigma}{\mathcal{P}(\text{Locks})}
      \category[Clauses]{c}{\Clause{\LockSetSigma}{\A}}
      \categoryFromSet[Policies]{\PolicyP}{\mathcal{P}(\text{Clauses})}
      \category[Types]{\AType} \alternative{\Unit} \alternative{\Nat} \alternative{\SumType{\TypeTauI}{\TypeTauII}}
      \alternative{\ProdType{\TypeTauI}{\TypeTauII} } \alternative{\RefType{\TypeTau}} \alternative{\ArrowType{\LockSetSigma}{\PolicyP}{\TypeTauI}{\TypeTauII}}
      \category[Annotated Types]{\TypeTau, \TypeTauI, \TypeTauII} \alternative{\AnnotateType{\AType}{\PolicyP}}
      \categoryFromSet[Variables]{x}{\mathbb{V}}
      \categoryFromSet[Numbers]{n}{\mathbb{N}}
      \categoryFromSet[Locations]{l}{\mathcal{L}}
      \category[Expressions]{e,e',e''} \alternative{x} \alternative{\UnitVal} \alternative{n} \alternative{\Lam{x}{e}} \alternative{\Pair{e}{e'}} \alternative{\Fst(e)} \alternative{\Snd(e)}
      \alternativeLine{\Inl(e)} \alternative{\Inr(e)} \alternative{\Case{e}{x}{e'}{y}{e''}}
      \alternativeLine{\App{e}{e'}} \alternative{\New(e,\TypeTau)}\alternative{\Deref{e}} \alternative{\Assgn{e}{e'}}
      \alternativeLine{\Open{\LockSigma}{e}} \addedmaths{\alternative{\Opened{\LockSigma}{e}}}
      \alternativeLine{\Close{\LockSigma}{e}} \addedmaths{\alternative{\Closed{\LockSigma}{e}}}
      \addedmaths{\alternative{\When{\LockSigma}{e}{e'}}}
      \alternative{l} 
      \category[Values~$\mathcal{V}$]{v,v'} \alternative{\Lam{x}{e}} \alternative{\Pair{v}{v'}} \alternative{\Fst(v)} \alternative{\Snd(v)} \alternative{\Inl(v)} \alternative{\Inr(v)} \alternative{\UnitVal} \alternative{l} \alternative{n}
    \end{syntax}
    \added{where $\mathbb{V}$ is a countably infinite set of variables and $\mathcal{L}$ is a countably infinite set of memory locations.}
  }
  \caption{Syntax}
  \label{syntax}
\end{figure*}

We base {\calcname} on FG~(\cite{RajaniG18}), a simply typed $\lambda$-calculus with a unit type, natural numbers, sums, pairs, references, functions, and information-flow labels.
We extend this core calculus with the declassification mechanism and the policy language of Flow~Locks~(\cite{BrobergS09,BrobergS06}).
\added{Additionally, we add the possibility of branching on locks and for changes to locks to be} \added{observed, based on an extension to Flow Locks known as Paralocks} \added{(\mbox{\cite{BrobergS10}})}.
The syntax of \del{this adapted language}\added{\calcname} is shown in Figure~\ref{syntax}.
\del{The accompanying technical appendix adds the possibility of branching on locks and for changes to locks to be observed, based on an extension to Flow Locks known as Paralocks \mbox{(\cite{BrobergS10}).}}
Note that, while our language does not explicitly include booleans and if expressions, we will use them freely in examples via their usual encoding as sum types.

Flow~Locks allows declassification via \emph{locks}---usually denoted by $\LockSigma, \LockSigmaPrime, \LockSigmaI$, etc.---which can be opened and closed using the \OpenKW and \CloseKW constructs.
We can think of these locks as boolean flags, denoting whether data is protected or not.
In other words, locks placed on data, while closed, prevent certain actors from seeing that data.
Opening the lock removes this protection.
Note that, unlike Flow Locks, {\calcname} uses syntactically scoped \OpenKW and \CloseKW statements.
This matches the functional nature of {\calcname} better than the unscoped, imperative setting of prior work.

To see how locks allow for declassification, consider the following example:
a conference wants to publish the list of accepted papers, but only after the program~committee has made all acceptance decisions.
The conference software contains the following function, which publishes a paper to a publicly visible output \Out{}\footnote{While a more realistic model would explicitly model input-output channels, we choose to minimize our calculus and use references for these purposes.
	This does not change the information-flow reasoning in any way.}:
$$\Publish \triangleq \Lam{x}{\Assgn{\Out}{x}}$$

The list of accepted papers could then be published by
$$\PublishAll \triangleq \App{\App{\LApply}{\Publish}}{\AcceptedPapers}$$
where $\AcceptedPapers$ is the list of accepted papers\added{ and $\LApply$ is a function corresponding to OCaml's \emph{iter} function, applying a function with side effects to every element of a list}.
While this correctly publishes the list of papers, it does not enforce our security policy, since nothing prevents this function from being run before the program~committee has made its decision.

To prevent this leak, we introduce a lock $\Published$ that protects papers from being published.
Thus, we associate the Flow-Locks \emph{policy}~$\Policy{\Clause{\{\Published\}}{\Pub}}$ with each paper, where \Pub is an \emph{actor} standing for the public.
This association tells us that the public is only allowed to see any information derived from a paper if the lock $\Published$ has been opened.
In general, we associate each term with a Flow-Locks \emph{policy} determining an information-flow policy.
These \emph{policies} are defined as sets of \emph{clauses} of the form $\Clause{\LockSetSigma}{\A}$, where $\LockSetSigma$ is a set of locks and $\A$ is an actor~(\cite{BrobergS06,BrobergS09}).
Information labeled with a clause $\Clause{\LockSetSigma}{\A}$ may be seen by the actor~$\A$ if all locks in $\LockSetSigma$ are open.
A policy allows an information flow if it is allowed by any clause.

Since \del{any data written to }\added{we want }\Out{} \added{to model an output}\del{can be observed by} the public\added{ can observe}, we associate the data stored in \Out{} with the policy $\Policy{\Clause{\EmptyLock}{\Pub}}$.
With this association, we must ensure that whenever we use \added{the }\Publish \added{function}, the \Published lock is statically known to be open.
To do so, we can change \PublishAll to check that the current date (stored in the variable \DateVar) is later than the publication date (stored in \PubDate), and then open the $\Published$~lock.
$$
\DefMacro{\PublishAll}
{
  \ExprIf*{\DateVar > \PubDate}
  {\Open*{\Published}{\App{\App{\LApply}{\Publish}}{\AcceptedPapers}}}
  {()}
}
$$

Alternatively, we could change \Publish to check the policy and open the lock if appropriate.
Remember that the opened lock is closed at the end of the scope of the \OpenKW expression.
Hence this approach opens (and closes) the lock for each assignment to \Out{} separately.
$$
\DefMacro{\Publish}
{
  \Lam*{x}
  {
    \ExprIf*{\DateVar > \PubDate}
    {\Open*{\Published}{\Assgn{\Out}{x}}}
    {()}
  }
}
$$
\added{
  Finally, we can leave it to the context of \mbox{\PublishAll} to set the lock, rather than change it ourselves.
  To do so, we adjust the definition of \mbox{\Publish} as follows (keeping the definition of \mbox{\PublishAll} the same):
}
$$
\addedmaths{\DefMacro{\Publish}
  {
    \Lam*{x}
    {
      \When*{\Published}
      {\Assgn{\Out}{x}}
      {()}
    }
  }}
$$
\added{
  This version would only release its input to the public if the lock $\Published$ is open.}

\added{Note that, because we can branch on and observe changes to locks, locks are relevant to information-flow reasoning.
  In other words, they can influence if and where a secret is revealed.
  Changes to locks can, in fact, reveal secrets in and of themselves.
  To take this into account, we follow \mbox{\cite{BrobergS10}} and also associate a policy $\LockPolicy(\LockSigma)$ with every lock $\LockSigma$.
  In the examples above, we consider all locks to have the least-restrictive policy $\PolicyBot \triangleq \Policy{\Clause{\EmptyLock}{\A} \mid \text{$\A$ is an actor}}$.
}

We will see in Section~\ref{sec:type-system} that each implementation of this example gets a different type.
We will also see how the type system uses those types to force the programmer to use the different~functions~correctly.

\subsection{Operational semantics}
\label{sec:oper-semant}
\begin{figure*}[t]
  { \small
    \begin{mathparpagebreakable}
      \addedmaths{\axiom{Eopen}
        {\Step{\Open{\LockSigma}{e},\LockSetSigma,S,\Opened{\LockSigma}{e},S,\OpenObs(\LockSigma),\LockSetSigma}}}
      \and
      % \del{\Infer{Eopened}
      % {\Step{e,\LockSetSigma \cup \set{\LockSigma},S,e',S',\ObservationOmega,\LockSetSigmaPrime}}
      % {\Step{\Open{\LockSigma}{e},\LockSetSigma,S,\Open{\LockSigma}{e'},S',\ObservationOmega,\LockSetSigmaPrime}}}
      \addedmaths{\Infer{Eopened}{\Step{e,\LockSetSigma \cup \set{\LockSigma},S,e',S',\ObservationOmega,\LockSetSigmaPrime}}{\Step{\Opened{\LockSigma}{e},\LockSetSigma,S,\Opened{\LockSigma}{e'},S',\ObservationOmega,\LockSetSigmaPrime}}}
      \and
      \addedmaths{\axiom{EopenedBeta}{\Step{\Opened{\LockSigma}{v},\LockSetSigma,S,v,S,\UnopenObs(\LockSigma),\LockSetSigma}}}
      % \Infer{EopenedBeta}{ }
      % {\Step{\Open{\LockSigma}{v},\LockSetSigma,S,v,S,\EmptyObs,\LockSetSigma}}
      \and
      \addedmaths{\axiom{Eclose}
        % {\sigma \in \Sigma}
        {\Step{\Close{\LockSigma}{e},\LockSetSigma,S,\Closed{\LockSigma}{e},S,\CloseObs(\LockSigma),\LockSetSigma}}}
      \and	
      \addedmaths{	\Infer{Eclosed}{\Step{e,\LockSetSigma \backslash \set{\LockSigma},S,e',S',\ObservationOmega,\LockSetSigmaPrime}}{\Step{\Closed{\LockSigma}{e},\LockSetSigma,S,\Closed{\LockSigma}{e'},S',\ObservationOmega,\LockSetSigmaPrime}}}
      % \Infer{Eclosed}
      % {\Step{e,\LockSetSigma \backslash \set{\LockSigma},S,e',S',\ObservationOmega,\LockSetSigmaPrime}}
      % {\Step{\Close{\LockSigma}{e},\LockSetSigma,S,\Close{\LockSigma}{e'},S',\ObservationOmega,\LockSetSigmaPrime}}
      \and
      \addedmaths{\axiom{EclosedBeta}{\Step{\Closed{\LockSigma}{v},\LockSetSigma,S,v,S,\UncloseObs(\LockSigma),\LockSetSigma}}}
      % \Infer{EclosedBeta}{ }
      % {\Step{\Close{\LockSigma}{v},\LockSetSigma,S,v,S,\EmptyObs,\LockSetSigma}}
      \and
      \Infer{EDerefBeta}{\InHeap{l}{v,\TypeTau}{S}}{\Step{\Deref{l},\LockSetSigma,S,v,S, \EmptyObs, \LockSetSigma}} \and
      \addedmaths{\Infer{ENew}{\Step{e,\LockSetSigma,S,e',S',\ObservationOmega,\LockSetSigmaPrime}}{\Step{\New{(e,\TypeTau)} ,\LockSetSigma,S,\New{(e',\TypeTau)} ,S', \ObservationOmega, \LockSetSigmaPrime}}} \and	
      \Infer{ENewBeta}
      {l \notin \Dom(S)}
      {\Step{\New{(v,\TypeTau)},\LockSetSigma,S,l,S \cup \{\HeapMapsTo{l}{v,\TypeTau}\},\WriteObs{l}{\TypeTau}(v),\LockSetSigma}}
      \and		
      \addedmaths{\Infer{Eassignl}{\Step{e,\LockSetSigma,S,e'',S',\ObservationOmega,\LockSetSigmaPrime}}{\Step{\Assgn{e}{e'},\LockSetSigma,S,\Assgn{e''}{e'},S',\ObservationOmega,\LockSetSigmaPrime}}}\and
      \addedmaths{\Infer{Eassignr}{\Step{e,\LockSetSigma,S,e',S',\ObservationOmega,\LockSetSigmaPrime}}{\Step{\Assgn{l}{e},\LockSetSigma,S,\Assgn{l}{e'},S',\ObservationOmega,\LockSetSigmaPrime}}}
      \and
      \Infer{Eassign}{l \in dom(S) \\ \HeapTypeLookup(S,l) = \TypeTau}
      {\Step{\Assgn{l}{v},\LockSetSigma,S,\UnitVal,S[\HeapMapsTo{l}{v,\TypeTau}],\WriteObs{l}{\TypeTau}(v),\LockSetSigma}} \and
      \addedmaths{
        \Infer{EAppl}{\Step{e,\LockSetSigma,S,e'',S',\ObservationOmega,\LockSetSigmaPrime}}{\Step{\App{e}{e'},\LockSetSigma,S,\App{e''}{e'},S',\ObservationOmega,\LockSetSigmaPrime}}}  \and
      \addedmaths{
        \Infer{EAppr}{\Step{e',\LockSetSigma,S,e'',S',\ObservationOmega,\LockSetSigmaPrime}}{\Step{\App{(\Lam{x}{e})}{e'},\LockSetSigma,S,\App{(\Lam{x}{e})}{e''},S',\ObservationOmega,\LockSetSigmaPrime}}} \and
      \addedmaths{
        \axiom{EAppBeta}{\Step{\App{(\Lam{x}{e})}{v},\LockSetSigma,S,[v/x]e,S, \EmptyObs, \LockSetSigma}}}
    \end{mathparpagebreakable}
  }
  \caption{Small-step reduction\del{ (selected rules)}}
  \label{reduction}
\end{figure*}
\begin{figure*}
  \ContinuedFloat
  {\small
    \begin{mathparpagebreakable}
      
       \addedmaths{\Infer{EPairl}{\Step{e,\LockSetSigma,S,e'',S',\ObservationOmega,\LockSetSigmaPrime}}{\Step{\Pair{e}{e'},\LockSetSigma,S,\Pair{e''}{e'},S',\ObservationOmega,\LockSetSigmaPrime}}}
      \and
      \addedmaths{\Infer{EPairr}{\Step{e,\LockSetSigma,S,e',S',\ObservationOmega,\LockSetSigmaPrime}}{\Step{\Pair{v}{e},\LockSetSigma,S,\Pair{v}{e'},S',\ObservationOmega,\LockSetSigmaPrime}}} \and
      \addedmaths{\Infer{EFst}{\Step{e,\LockSetSigma,S,e',S',\ObservationOmega,\LockSetSigmaPrime}}{\Step{\Fst{(e)},\LockSetSigma,S,\Fst{(e')},S',\ObservationOmega,\LockSetSigmaPrime}}} \and
      
      \addedmaths{\axiom{EFstBeta}{\Step{\Fst{(\Pair{v}{v'})},\LockSetSigma,S,v,S, \EmptyObs, \LockSetSigma}}}
      \and
      \addedmaths{\Infer{ESnd}{\Step{e,\LockSetSigma,S,e',S',\ObservationOmega,\LockSetSigmaPrime}}{\Step{\Snd{(e)},\LockSetSigma,S,\Snd{(e')},S',\ObservationOmega,\LockSetSigmaPrime}}} \and
      \addedmaths{\axiom{ESndBeta}{\Step{\Snd{(\Pair{v}{v'})},\LockSetSigma,S,v',S, \EmptyObs, \LockSetSigma}}}\and
      \addedmaths{\Infer{EInl}{\Step{e,\LockSetSigma,S,e',S',\ObservationOmega,\LockSetSigmaPrime}}{\Step{\Inl{(e)},\LockSetSigma,S,\Inl{(e')},S',\ObservationOmega,\LockSetSigmaPrime}}} \and
      \addedmaths{\Infer{EInr}{\Step{e,\LockSetSigma,S,e',S',\ObservationOmega,\LockSetSigmaPrime}}{\Step{\Inr{(e)},\LockSetSigma,S,\Inr{(e')},S',\ObservationOmega,\LockSetSigmaPrime}}} \and

      \addedmaths{\Infer{ECase}{\Step{e,\LockSetSigma,S,e''',S',\ObservationOmega,\LockSetSigmaPrime}}{\Step{\Case{e}{x}{e'}{y}{e''},\LockSetSigma,S,\Case{e'''}{x}{e'}{y}{e''},S',\ObservationOmega,\LockSetSigmaPrime}}} \and
      \addedmaths{\axiom{ECasel}{\Step{\Case{\Inl{(v)}}{x}{e'}{y}{e''},\LockSetSigma,S,[v/x]e',S, \EmptyObs, \LockSetSigma}}} \and
      \addedmaths{\axiom{ECaser}{\Step{\Case{\Inr{(v)}}{x}{e'}{y}{e''},\LockSetSigma,S,[v/x]e'',S, \EmptyObs, \LockSetSigma}}}
      \and
      \addedmaths{\Infer{EWhenOpen}{\LockSigma \in \LockSetSigma\\ \Step{e,\LockSetSigma,S,e'',S',\ObservationOmega,\LockSetSigmaPrime}}
        {\Step{\When{\LockSigma}{e}{e'},\LockSetSigma,S,\When{\LockSigma}{e''}{e'},S',\ObservationOmega,\LockSetSigmaPrime}}} \and
      \addedmaths{\Infer{EWhenClosed}{\LockSigma \notin \LockSetSigma \\\Step{e',\LockSetSigma,S,e'',S',\ObservationOmega,\LockSetSigmaPrime}}
        {\Step{\When{\LockSigma}{e}{e'},\LockSetSigma,S,\When{\LockSigma}{e}{e''},S',\ObservationOmega,\LockSetSigmaPrime}}} \and
      \addedmaths{\Infer{EWhenOpenBeta}{\LockSigma \in \LockSetSigma }{\Step{\When{\LockSigma}{v}{e'},\LockSetSigma,S,v,S,\EmptyObs,\LockSetSigma}}}  \and
      \addedmaths{\Infer{EWhenClosedBeta}{\LockSigma \notin \LockSetSigma}{\Step{\When{\LockSigma}{e}{v},\LockSetSigma,S,v,S,\EmptyObs,\LockSetSigma}}}
    \end{mathparpagebreakable}
  }
  \caption{\added{Small-step reduction (continued)}}

\end{figure*}

We use a small-step, call-by-value operational semantics.
In order to define this semantics we need some knowledge about the \emph{context} in which programs execute.
The context contains a (heap) state~$S$ which maps finitely many locations to both the value currently stored at that location and the type of that value.
If $\InHeap{l}{v,\TypeTau}{S}$ for a state~$S$, we say that $\HeapLookup{S}(l) = v$ and $\HeapTypeLookup(S,l) = \TypeTau$.
Because execution steps can change the state, every execution step results in a new state in addition to the reduced~term.

Flow Locks' attacker model~(\cite{BrobergS09}), which we use, assumes that an attacker can observe changes to the state.
To model this assumption, each reduction step also outputs an \emph{observation}~$\ObservationOmega$, which specifies what changes to the state an attacker can see due to the step.
There are two types of observations \added{we take}\del{(also} from Flow Locks~(\cite{BrobergS09,BrobergS10})\del{)}: the empty observation~$\EmptyObs$ and~$\WriteObs{l}{\TypeTau}(v)$, which says that the value~$v$  was written to location~$l$ that should hold values of type $\TypeTau$.
\added{Additionally, we add the $\OpenObs(\LockSigma)$ and $\CloseObs(\LockSigma)$ observations from Paralocks (\mbox{\cite{BrobergS10}}), which witness the explicit opening and closing of a lock $\LockSigma$.
Due to the scoped nature of our opening and closing constructs, we also add two observations $\UnopenObs(\LockSigma)$ and $\UncloseObs(\LockSigma)$, which witness the end of the scope of an \mbox{\OpenKW} or \mbox{\CloseKW} operation, respectively.}
We define the policy of an observation as follows:
\[
  \begin{array}{lll}
    \ObsPolicy(\ObservationOmega) & = &
                                        \left\lbrace
                                        \begin{array}{ll}
                                          \PolicyP & \mbox{if } \ObservationOmega = \WriteObs{l}{\TypeTau}(v) \mbox{ and } \TypeTau = \AnnotateType{\AType}{\PolicyP} \\
                                          \addedmaths{\LockPolicy(\LockSigma)} & \addedmaths{\mbox{if } \ObservationOmega = \OpenObs(\LockSigma) \mbox{ or } \ObservationOmega = \CloseObs(\LockSigma)}\\
                                          \addedmaths{\LockPolicy(\LockSigma)} & \addedmaths{\mbox{if } \ObservationOmega = \UnopenObs(\LockSigma) \mbox{ or } \ObservationOmega = \UncloseObs(\LockSigma)}\\
                                          \mbox{undefined} & \mbox{if } \ObservationOmega = \EmptyObs
                                        \end{array}
                                        \right.
  \end{array}
\]

Our operational semantics also tracks which locks are open when a step is taken.
\del{Since the language we present here does }
\added{This would be necessary even if we did }not allow executions to branch depending on the lock state\del{.}\del{this may be surprising.
  However}, \added{since }these locks are used in our definition of security\added{~(Section \mbox{\ref{sec:comp-defin-secur}})}.
\del{Moreover, we allow branching on the lock state in the extended language of the technical appendix.}

One might expect that we return a new lock set after every step, just as we do for the state.
In particular, one might na\"{i}vely expect $\Open{\LockSigma}{e}$ to reduce to $e$ with a new lock set \mbox{$\LockSetSigma \cup \set{\LockSigma}$} when executed for one step in $\LockSetSigma$.
In fact, this is exactly how \citet{BrobergS06,BrobergS09,BrobergS10} handle changes to the lock state.
However, because in our setting \OpenKW and \CloseKW are scoped, we need to return to the original lock set after their scope has ended.
If we use the na\"{i}ve semantics, we lose vital information.
To see this, consider the three programs \mbox{$\Open{\LockSigma}{(\App{e}{e'})}$}, $\App{(\Open{\LockSigma}{e})}{e'}$, and $\App{(\Open{\LockSigmaPrime}{e})}{e'}$.
Assume that the first two programs are executed in lock set $\set{\LockSigmaPrime}$ and the last one is executed in $\set{\LockSigma}$.
With the \added{na\"{i}ve }semantics above all of them would reduce in one step to $\App{e}{e'}$ with a new lock set $\set{\LockSigma,\LockSigmaPrime}$.
We would thus lose the ability to distinguish them, so we would not know when the scope of the \OpenKW ends and which locks to reset.
To solve this problem, we reduce $e$ \emph{inside} the scope of the \OpenKW{}.
The lock $\LockSigma$ is open within this scope, i.e., while $e$ reduces.
When $e$ has reduced to a value~$v$, we remove the \OpenKW{} and return $v$.\added{\footnote{\added{This is a slight simplification, which we will correct when we discuss the relevant reduction rules.}}}

Next, consider the program $\Open{\LockSigma}{\New(\Const{5},\AnnotateType{\Nat}{\PolicyP})}$ with open locks~$\LockSetSigma$.
The security of this program depends on the security of the subterm $\New(\Const{5},\AnnotateType{\Nat}{\PolicyP})$ with open locks~$\LockSetSigma \cup \{\LockSigma\}$.
Since this subterm is the redex of the full term, we say that $\LockSetSigma \cup \{\LockSigma\}$ is the \emph{active} lock set in the reduction of the full term.
In general, the active lock set of a reduction step is the lock set at the redex reduced by the step.
We need to know this active lock set in order to determine whether the reduction step is secure.
We therefore instrument our small-step relation to explicitly track the active lock set.

We now have the pieces to define reduction.
A reduction step has the form \smash{$\Step{e,\LockSetSigma,S,e',S',\ObservationOmega,\LockSetSigmaPrime}$}, which means that the expression~$e$, in set of open locks~$\LockSetSigma$ and state~$S$, can take a step to the expression~$e'$, changing the state to $S'$ and producing an observation~$\ObservationOmega$.
The lock set $\LockSetSigmaPrime$ is the active lock set at the point of reduction.
\del{Selected}\added{The} reduction rules can be found in Figure~\ref{reduction}.
\del{The full rules can be found in our technical appendix.}

The reduction rules define a \added{left-to-right, }call-by-value semantics.
The initial lock set is static and remains constant over several steps of execution, while the structure of the expression being executed determines the \del{effective}\added{active} lock set at the point of reduction.
Most rules are standard, apart from the lock set and observations.
Most $\beta$-reduction steps are unobservable, and thus produce the observation~$\EmptyObs$.
One example is \rname{EDerefBeta} which reads a value stored in an existing location and silently outputs~$\EmptyObs$.
\added{In the following, we discuss only those rules that are non-standard or interesting from an information-flow~perspective.}

Memory-manipulating reductions produce nontrivial observations.
The rule \rname{ENewBeta} allocates a new location~$l$, storing the value~$\Const{v}$ and the specified type~$\TypeTau$ in $l$.
It produces the observation $\WriteObs{l}{\TypeTau}(\Const{v})$ and returns the new location~$l$.
\rname{Eassign} updates an existing location~$l$ of type $\TypeTau$ in the state $S$ to a value~$\Const{v}$ (written $S[\HeapMapsTo{l}{v,\TypeTau}]$) and reduces to~$\UnitVal$ with observation~$\WriteObs{l}{\TypeTau}(\Const{v})$.

Finally, the lock-manipulating rules determine our declassification mechanism.
\added{The rule $\rname{Eopen}$ reduces $\Open{\LockSigma}{e}$ to the runtime term $\Opened{\LockSigma}{e}$ and produces the observation $\OpenObs(\LockSigma)$.
  This change from $\Open{\LockSigma}{e}$ to $\Opened{\LockSigma}{e}$ allows us to syntactically represent the fact that the $\OpenObs(\LockSigma)$ observation has already been produced, while preserving the information that the scope of the \mbox{\OpenKW} is restricted to expression $e$.}

The rule \rname{Eopened} allows $e$ to reduce with the lock~$\LockSigma$ open \added{within the scope of the \mbox{\OpenedKW} construct}.
Once $e$ has reduced to a value~$v$, \rname{EopenedBeta} returns $v$ with observation \del{$\EmptyObs$}\addedmaths{\OpenObs(\LockSigma)}.
The rules for \CloseKW work similarly.

\added{The rules for the \mbox{\WhenKW} construct reduce the expression inside the left or the right branch of the \mbox{\WhenKW} depending on the value of the lock (rules \mbox{\rname{EWhenOpen}} and \mbox{\rname{EWhenClosed}}).}
\added{Once the branch under consideration has evaluated to a value, the rules \mbox{\rname{EWhenOpenBeta}} and \mbox{\rname{EWhenClosedBeta}} reduce the whole \mbox{\WhenKW} construct to that value.}

\added{The choice to reduce the chosen branch of a \mbox{\WhenKW} construct \emph{inside} its parent (rather than stepping directly to that branch) may seem odd.
We choose this reduction strategy because with it the resulting program after the reduction step still contains both branches and can therefore still be used safely in lock sets both with and without the branched-on lock being open.
}
\added{As a result, we can state and prove some properties without consideration of what lock set we are in, making those proofs easier.
}

% !TeX spellcheck = en_US
\section{Type system}
\label{sec:type-system}

\begin{figure*}
{ \small
  \begin{mathpar}
    \Infer{Sub-Policy}
    {
      \PolicyOrder{\PolicyP}{\PolicyPPrime}\\
      \Subtyping{\AType}{\BType}
    }
    {
      \Subtyping{\AnnotateType{\AType}{\PolicyP}}{\AnnotateType{\BType}{\PolicyPPrime}}
    } 
\and
\addedmaths{\axiom{sub-ref}{\Subtyping{\RefType{\TypeTau}}{\RefType{\TypeTau}}}}
 \and
\addedmaths{\Infer{sub-prod}{\Subtyping{\TypeTauN}{\TypeTauI} \and \Subtyping{\TypeTauII}{\TypeTauIII}}{\Subtyping{\ProdType{\TypeTauN}{\TypeTauII}}{\ProdType{\TypeTauI}{\TypeTauIII}}}} \and
\addedmaths{\Infer{sub-sum}{\Subtyping{\TypeTauN}{\TypeTauI} \\ \Subtyping{\TypeTauII}{\TypeTauIII}}{\Subtyping{\SumType{\TypeTauN}{\TypeTauII}}{\SumType{\TypeTauI}{\TypeTauIII}}}}
\and
    \Infer{Sub-Arrow}
    {
      \Subtyping{\TypeTauN}{\TypeTauI}\\ \Subtyping{\TypeTauII}{\TypeTauIII} \\ \PolicyPPrime \sqsubseteq \PolicyP \\ \newstuff{\LockSetSigma \subseteq \LockSetSigmaPrime}
    }
    {
      \Subtyping{\ArrowType{\newstuff{\LockSetSigma}}{\PolicyP}{\TypeTauI}{\TypeTauII}}{\ArrowType{\newstuff{\LockSetSigmaPrime}}{\PolicyPPrime}{\TypeTauN}{\TypeTauIII}}
    } \and
\addedmaths{\axiom{sub-unit}{\Subtyping{\Unit}{\Unit}}} \and
\addedmaths{\axiom{sub-nat}{\Subtyping{\Nat}{\Nat}}}
\end{mathpar}
}
\caption{\added{Subtyping rules}}
\label{fig:subtyping-rules}
\end{figure*}
\begin{figure*}
	{ \small
\begin{mathparpagebreakable}
	\addedmaths{\axiom{var}{\Typing{\Gamma,\hyp{x}{\TypeTau},\Gamma';\newstuff{\LockSetSigma};\theta}{\pc}{x}{\TypeTau}}} \and
	\addedmaths{\axiom{unit}{\Typing{\Gamma;\newstuff{\LockSetSigma};\theta}{\pc}{\UnitVal}{\AnnotateType{\Unit}{\PolicyBot}}}} \and
	\addedmaths{\Infer{nat}{n \in \NN}{\Typing{\Gamma; \newstuff{\LockSetSigma}; \theta}{\pc}{n}{\AnnotateType{\Nat}{\PolicyBot}}}} \and
    \Infer{$\lambda$}
    {\Typing{\Gamma, \hyp{x}{\TypeTauI};\newstuff{\LockSetSigmaPrime};\theta}{\pcPrime}{e}{\TypeTauII}}
    {\Typing{\Gamma;\newstuff{\LockSetSigma};\theta}{\pc}{\Lam{x}{e}}{\AnnotateType{(
          \ArrowType{\newstuff{\LockSetSigma'}}{\pcPrime}{\TypeTauI}{\TypeTauII})}{\PolicyBot}}} \and
      \newstuff{
    \Infer{open}
    {
      \Typing{\Gamma; \LockSetSigma \cup \{\LockSigma\}; \theta}{\pc}{e}{\TypeTau} \and \addedmaths{\PolicyOrder{\pc}{\LockPolicy(\LockSigma)} }
    }
    {\Typing{\Gamma; \LockSetSigma; \theta}{\pc}{\Open{\sigma}{e}}{\TypeTau}}}
    \and
    \newstuff{\addedmaths{
    \Infer{opened}
    {
    	\Typing{\Gamma; \LockSetSigma \cup \{\LockSigma\}; \theta}{\pc}{e}{\TypeTau} \and \PolicyOrder{\pc}{\LockPolicy(\LockSigma)}
    }
    {\Typing{\Gamma; \LockSetSigma; \theta}{\pc}{\Opened{\sigma}{e}}{\TypeTau}}}}
\and
    \newstuff{
    \Infer{close}
    {
      \Typing{\Gamma;\LockSetSigma \setminus \{\LockSigma\}; \theta}{\pc}{e}{\TypeTau} \and \addedmaths{\PolicyOrder{\pc}{\LockPolicy(\LockSigma)} }
    }
    {\Typing{\Gamma;\LockSetSigma;\theta}{\pc}{\Close{\sigma}{e}}{\TypeTau}}}
\and
\newstuff{\addedmaths{
		\Infer{closed}
		{
			\Typing{\Gamma; \LockSetSigma \setminus \{\LockSigma\}; \theta}{\pc}{e}{\TypeTau} \and \PolicyOrder{\pc}{\LockPolicy(\LockSigma)}
		}
		{\Typing{\Gamma; \LockSetSigma; \theta}{\pc}{\Closed{\sigma}{e}}{\TypeTau}}}}
    \and
    \addedmaths{
    	\Infer{prod}
    	{
    		\Typing{\Gamma; \newstuff{\LockSetSigma}; \theta}{\pc}{e_1}{\TypeTauI} \and
    		\Typing{\Gamma; \newstuff{\LockSetSigma}; \theta}{\pc}{e_2}{\TypeTauII}
    	}
    	{
    		\Typing{\Gamma; \newstuff{\LockSetSigma}; \theta}{\pc}{\Pair{e_1}{e_2}}{\AnnotateType{(\ProdType{\TypeTauI}{\TypeTauII})}{\PolicyBot}}
    	}
    
} \and
\addedmaths{
	\Infer{fst}{\Typing{\Gamma; \newstuff{\LockSetSigma}; \theta}{\pc}{e}{\AnnotateType{(\ProdType{\TypeTauI}{\TypeTauII})}{\PolicyP}} \and \PolicyOrder{\PolicyP}{\TypeTauI}}{\Typing{\Gamma; \newstuff{\LockSetSigma}; \theta}{\pc}{\Fst(e)}{\TypeTauI}}
}
\and
\addedmaths{
	\Infer{snd}{\Typing{\Gamma; \newstuff{\LockSetSigma}; \theta}{\pc}{e}{\AnnotateType{(\ProdType{\TypeTauI}{\TypeTauII})}{\PolicyP}} \and \PolicyOrder{\PolicyP}{\TypeTauII}}{\Typing{\Gamma; \newstuff{\LockSetSigma}; \theta}{\pc}{\Snd(e)}{\TypeTauII}}
} \and 
\addedmaths{
	\Infer{inl}{\Typing{\Gamma; \newstuff{\LockSetSigma}; \theta}{\pc}{e}{\TypeTauI}}{\Typing{\Gamma; \newstuff{\LockSetSigma}; \theta}{\pc}{\Inl(e)}{\AnnotateType{(\SumType{\TypeTauI}{\TypeTauII})}{\PolicyBot}}}
}
\and 
\addedmaths{
	\Infer{inr}{\Typing{\Gamma; \newstuff{\LockSetSigma}; \theta}{\pc}{e}{\TypeTauII}}{\Typing{\Gamma; \newstuff{\LockSetSigma}; \theta}{\pc}{\Inr(e)}{\AnnotateType{(\SumType{\TypeTauI}{\TypeTauII})}{\PolicyBot}}}
}
\and
    \Infer{case}
    {\Typing{\Gamma;\newstuff{\LockSetSigma};\theta}{\pc}{e}{\AnnotateType{(\SumType{\TypeTauI}{\TypeTauII})}{\PolicyP}} \\ \PolicyOrder{\PolicyP}{\TypeTau} \\ \Typing{\Gamma,\hyp{x}{\TypeTauI};\newstuff{\LockSetSigma};\theta}{\PolicyJoin{\pc}{\PolicyP}}{e_1}{\TypeTau} \\ \Typing{\Gamma,\hyp{y}{\TypeTauII};\newstuff{\LockSetSigma};\theta}{\PolicyJoin{\pc}{\PolicyP}}{e_2}{\TypeTau}
    }
    {\Typing{\Gamma;\newstuff{\LockSetSigma};\theta}{\pc}{\Case{e}{x}{e_1}{y}{e_2}}{\TypeTau}}
    \and
\addedmaths{\Infer{sub}
	{\Typing{\Gamma;\newstuff{\LockSetSigma};\theta}{\pcPrime}{e}{\TauPrime} \and \PolicyOrder{\pc}{\pcPrime} \and \Subtyping{\TauPrime}{\TypeTau}}
	{\Typing{\Gamma;\newstuff{\LockSetSigma};\theta}{\pc}{e}{\TypeTau}}
	}    
    \and
    \Infer{app}
    {
      \Typing{\Gamma;\newstuff{\LockSetSigma};\theta}{\pc}{e_1}{\AnnotateType{(\ArrowType{\newstuff{\LockSetSigmaPrime}}{\pcPrime}{\TypeTauI}{\TypeTauII})}{\PolicyP}}\\
      \Typing{\Gamma;\newstuff{\LockSetSigma};\theta}{\pc}{e_2}{\TauPrimeI}\\
      \PolicyOrder{\PolicyP}{\TypeTauII}\\
      \PolicyOrder{\PolicyJoin{\pc}{\PolicyP}}{\pcPrime}\\
      \Subtyping{\TauPrimeI}{\TypeTauI}\\
      \newstuff{\LockSetSigma \supseteq \LockSetSigmaPrime}
    }{\Typing{\Gamma;\LockSetSigma;\theta}{\pc}{\App{e_1}{e_2}}{\TypeTauII}}
      \end{mathparpagebreakable}
}
\caption{\del{Subtyping and }Type system\del{ (selected rules)}}
\label{typing-rules}
\end{figure*}

\begin{figure*}
	\ContinuedFloat
	{\small
		\begin{mathparpagebreakable}
    \Infer{deref}
    {
      \Typing{\Gamma;\newstuff{\LockSetSigma};\theta}{\pc}{e}{\AnnotateType{(\RefType{\TypeTau})}{\PolicyP}}\\
      \PolicyOrder{\PolicyP}{\TauPrime}\\
      \Subtyping{\TypeTau}{\TauPrime}
    }
    {\Typing{\Gamma;\newstuff{\LockSetSigma};\theta}{\pc}{\Deref{e}}{\TauPrime}}
    \and
    \Infer{new}
    {
      \Typing{\Gamma;\newstuff{\LockSetSigma};\theta}{\pc}{e}{\TauPrime}\\
      \PolicyOrder{\pc}{\TypeTau}\\
      \Subtyping{\newstuff{\PolicySpecialize{\TauPrime}{\LockSetSigma}}}{\TypeTau}
    }
    {\Typing{\Gamma;\newstuff{\LockSetSigma};\theta}{\pc}{\New(e,\TypeTau)}{\AnnotateType{(\RefType{\TypeTau})}{\PolicyBot}}}
    \and
    \Infer{loc}{ }{\Typing{\Gamma;\newstuff{\LockSetSigma}; \theta}{\pc}{l}{\AnnotateType{(\RefType{\theta(l)})}{\PolicyBot}}}
    \and
    \Infer{assign}
    {
      \Typing{\Gamma;\newstuff{\LockSetSigma};\theta}{\pc}{e}{\AnnotateType{(\RefType{\TauPrime})}{\PolicyP}}\\
      \Subtyping{\newstuff{\PolicySpecialize{\TypeTau}{\LockSetSigma}}}{\TauPrime}\\
      \Typing{\Gamma;\LockSetSigma;\theta}{\pc}{e'}{\TypeTau}\\
      \PolicyOrder{\PolicyJoin{\pc}{\PolicyP}}{\TauPrime}
    }
    {\Typing{\Gamma;\newstuff{\LockSetSigma};\theta}{\pc}{\Assgn{e}{e'}}{\AnnotateType{\Unit}{\PolicyBot}}}
\and
\newstuff{\addedmaths{\Infer{when}{\Typing{\Gamma;\LockSetSigma \cup \LockSigma;\theta}{\PolicyJoin{\pc}{\LockPolicy(\LockSigma)}}{e_1}{\TypeTau} \and \Typing{\Gamma;\LockSetSigma;\theta}{\PolicyJoin{\pc}{\LockPolicy(\LockSigma)}}{e_2}{\TypeTau} \and \PolicyOrder{\LockPolicy(\LockSigma)}{\TypeTau}}
		{\Typing{\Gamma;\LockSetSigma;\theta}{\pc}{\When{\LockSigma}{e_1}{e_2}}{\TypeTau}}}}
  \end{mathparpagebreakable}
}
  \caption{\added{Type system (continued)}}
\end{figure*}

Our type system extends the FG~type system~(\cite{RajaniG18}), which enforces noninterference without any declassification.
Most rules remain unchanged, though we add rules for \OpenKW{}\added{,}\del{ and} \CloseKW{}\added{, and \mbox{\WhenKW}} and make small changes necessary for declassification to the rules manipulating state.
We annotate every type with a Flow-Locks policy~(\cite{BrobergS09,BrobergS06}), including types contained within compound~types.
\added{For example, in a pair type, we not only annotate the entire pair type with a policy, but each component of the pair gets its own policy as well.}

We start by reviewing the properties of Flow-Locks policies as explored by~\citet{BrobergS06,BrobergS09}.
As \del{is }usual, policies form a join semilattice.
A policy $\PolicyP$ is below a policy $\PolicyQ$, written $\PolicyOrder{\PolicyP}{\PolicyQ}$, if policy~$\PolicyP$ allows more actors to observe data labeled by it than does~$\PolicyQ$.
More specifically, this means that for every clause in $\PolicyQ$ there is a less-restrictive clause in $\PolicyP$.
A clause~$\Clause{\LockSetSigmaI}{\A}$ is less restrictive than~$\Clause{\LockSetSigmaII}{\A}$ if $\LockSetSigmaI \subseteq \LockSetSigmaII$.
The least-restrictive policy $\PolicyBot \triangleq \Policy{\Clause{\EmptyLock}{\A} \mid \text{$\A$ is an actor}}$ allows every actor to observe the labeled data in any lock state, while the most-restrictive policy $\PolicyTop \triangleq \EmptyPolicy$ does not allow any actor to observe the labeled data in any lock state.
The general principle of secure information flow is that\added{, in the absence of declassification,} information should only flow upwards in the policy lattice: data labeled $\PolicyP$ may flow to a location labeled $\PolicyQ$ if $\PolicyOrder{\PolicyP}{\PolicyQ}$.
For this reason, when $\PolicyOrder{\PolicyP}{\PolicyQ}$, we also say that policy~$\PolicyP$~\emph{flows to}~$\PolicyQ$.

The join of two policies $\PolicyP$ and $\PolicyQ$ is $\PolicyJoin{\PolicyP}{\PolicyQ} \triangleq \Policy{\Clause{\LockSetSigmaI \cup \LockSetSigmaII}{\A} \mid \Clause{\LockSetSigmaI}{\A} \in \PolicyP \land \Clause{\LockSetSigmaII}{\A} \in \PolicyQ}$.
Finally, policies can be updated, or \emph{specialized}, with regard to the current lock set.
The specialization of $\PolicyP$ with $\LockSetSigma$ (written~$\PolicySpecialize{\PolicyP}{\LockSetSigma}$) represents the ``effective'' policy $\PolicyP$ when locks in $\LockSetSigma$ are assumed to be open: $\PolicySpecialize{\PolicyP}{\LockSetSigma} \triangleq \Policy{\Clause{\LockSetSigmaI \backslash \LockSetSigma}{a} \mid \Clause{\LockSetSigmaI}{a} \in \PolicyP}$.

We lift the ordering on policies and policy specialization to policy-annotated types.
\begin{definition}
		\label{type-ordering}
		\begin{align*}			
		\PolicyOrder{\AnnotateType{\AType}{\PolicyP}}{\PolicyPPrime} &\triangleq \PolicyOrder{\PolicyP}{\PolicyPPrime} \\
		\PolicyOrder{\PolicyP}{\AnnotateType{\AType}{\PolicyPPrime}} &\triangleq \PolicyOrder{\PolicyP}{\PolicyPPrime} \\
		\PolicyOrder{\AnnotateType{\AType}{\PolicyP}}{\AnnotateType{\BType}{\PolicyPPrime}} &\triangleq\PolicyOrder{\PolicyP}{\PolicyPPrime}
		\end{align*}
		\centering
		
		For $\TypeTau = \AnnotateType{\AType}{\PolicyP}$ we define $\PolicySpecialize{\TypeTau}{\LockSetSigma} \triangleq \AnnotateType{\AType}{\PolicySpecialize{\PolicyP}{\LockSetSigma}}$.\\
\end{definition}
Typing judgments have the form $\Typing{\Gamma;\LockSetSigma;\theta}{\pc}{e}{\tau}$, where the context~$\Gamma$ maps variables to their annotated types, the lock set~$\LockSetSigma$ represents the open locks, the \emph{state environment}~$\theta$ gives types to locations, and \pc{} is a policy.
A judgment \mbox{$\Typing{\Gamma;\LockSetSigma;\theta}{\pc}{e}{\TypeTau}$} means that in environment $\Gamma$ expression $e$ has type $\TypeTau$ if at least the locks in $\LockSetSigma$ are open and locations store data of the types specified by $\theta$.
The policy $\pc$ is a lower bound on the policies of observations produced by $e$.

Our type system enjoys subtyping, which we write~$\TypeTauI \mathrel{<:} \TypeTauII$ meaning that $\TypeTauI$ is a subtype of~$\TypeTauII$.
As usual, this relationship implies that all terms of type $\TypeTauI$ can be used as if they had type $\TypeTauII$ (the restrictions on use posed by the policies in type $\TypeTauII$ are stricter than those posed by $\TypeTauI$).
\del{Selected}\added{The} typing and subtyping rules for our type system can be found in Figure\added{s \mbox{\ref{fig:subtyping-rules}} and}~\ref{typing-rules}\del{, with the full type system found in the accompanying technical~appendix}.
\del{The remaining subtyping rules are entirely standard and unchanged from FG~(\mbox{\cite{RajaniG18}}).}
We highlight changes from FG~(\cite{RajaniG18}) due to the addition of locks and declassification \added{in both the typing and subtyping rules }by giving the changes a \newstuff{\text{\hspace{-0.25em}yellow background\hspace{-0.25em}}}.

\added{Most subtyping rules are entirely standard and unchanged from FG~(\mbox{\cite{RajaniG18}}).}
\added{Base types are only subtypes of themselves.
Most other types require the constituent types to be subtypes of each other.
The rule \mbox{\rname{Sub-Policy}} additionally says that a type~$\TypeTau$ annotated with a (less-restrictive) policy~$\PolicyP$ is a subtype of the same type~$\TypeTau$ with a more-restrictive policy.
Reference types do not allow any non-trivial subtyping because they can be used in both read and write operations, which would require subtyping in opposite directions.
Instead, the typing rules for reading from and writing to references take subtyping into account themselves.
The only subtyping rule interacting with our declassification mechanism is \mbox{\rname{Sub-Arrow}} for function types,
which specifies that}
\del{For instance,} \added{(}in addition to standard subtyping requirements\added{)}\del{,} function types may require fewer locks than their supertypes.

\added{
For most typing rules, the only notable change from FG~(\mbox{\cite{RajaniG18}}) is that the context also contains a set of open locks $\LockSetSigma$.
Below, we will only explicitly discuss typing rules which are either new, interact with locks or declassification in some way, or are interesting from an information-flow perspective.

The subtyping rules discussed above are incorporated into the type system via the \mbox{\rname{Sub}} rule.
In addition to subtyping, it also allows the \pc~annotation to change.
Since the \pc~is an upper bound on the levels at which side effects may happen, a program with a higher \pc~annotation can also be used where only a lower annotation is required.}

The rules \rname{$\lambda$} and \rname{app} deal with functions.
In addition to annotating functions with a program-counter label~\pcPrime{}, we also annotate function types with a set of locks $\LockSetSigmaPrime$, representing the locks which must be open for the function body to run.
To ensure that the body can be safely run if the locks in $\LockSetSigmaPrime$ are open, the rule~\rname{$\lambda$} checks that the body type~checks with $\LockSetSigmaPrime$.

\added{Conversely, t}\del{T}he rule~\rname{app} ensures that a function call is safe by checking \del{that the local \pc{} flows to \pcPrime{}, and then checks }that the open locks at the point of application are a superset of $\LockSetSigmaPrime$
\added{and that the local \pc{} flows to \pcPrime{}}.
Moreover, \rname{app} checks that the policy $\PolicyP$ on the function type flows to the output of the function and to \pcPrime{}, ensuring that information about the function is not leaked via either the result or any effects in the function body.
This reasoning also applies to the rules \rname{deref} and \rname{assign}, which manipulate references, ensuring that the label $\PolicyP$ flows to the type of data read from or stored in a reference, respectively.

The rules \rname{open} and \rname{close} simply check that the body of the respective construct is well typed when the lock being opened or closed is added to or removed from the original lock set.
\added{The rules \mbox{\rname{opened}} and \mbox{\rname{closed}} do the same for the constructs \mbox{\OpenedKW} and \mbox{\ClosedKW}, since they represent an \mbox{\OpenKW} or \mbox{\CloseKW} after the first reduction step that produced the $\OpenObs(\LockSigma)$ or $\CloseObs(\LockSigma)$ observation.}

The rule \rname{case} ensures that the $\PolicyP$ labeling the data being case-analyzed is allowed to flow to the level of the result type.
Additionally it prevents leaks via implicit information flows by checking the two branches with a \pc~that is at least $\PolicyP$.

The typing rules \rname{new} and \rname{assign} for memory allocation and assignment allow for declassification by specialization with the current lock set.
Hence, the set of current locks is taken into account when writing to memory.
Note that this specialization is not done when checking for implicit information flows (e.g., in the rule for \CaseKW{} or function application).
To see why, consider typing the following expression in a lock set $\LockSetSigma$ which contains $\LockSigma$.
$$\Case*{x}{y}{\Close{\LockSigma}{\Assgn{l}{\Inl{\UnitVal}}}}{z}{\Close{\LockSigma}{\Assgn{l}{\Inr{\UnitVal}}}}$$

This expression leaks information about~$x$ to~$l$.
However, if we took the current lock set into account, we might allow branching on $x$ because $\LockSigma$ is open and, hence, $x$ is declassified.
But in the branches\added{, where the information is actually revealed,} $\LockSigma$ is closed, which could result in a leak that is not in compliance with the policy.

\added{The \mbox{\rname{When}} rule checks that both branches are typed at the same type.
	However, since the first branch is only executed if lock $\LockSigma$ is open, we explicitly add it to the set of open locks for type checking.
	We do not need to explicitly remove $\LockSigma$ from the lock set for the second branch because type checking is monotone in the lock set:
	any term that is well typed without a lock being open is also well typed with the lock being open.
	Similar to \mbox{\CaseKW} statements, information about \emph{which} branch of a \mbox{\WhenKW} executes can reveal the state of the lock that was branched on. Therefore, we also check that the policy of the lock is below the policy of the result type to ensure that information about the lock is not leaked in the result in violation of the policy.
	To ensure information about the lock is not leaked via side effects, we raise the \pc~annotation  when type checking the two branches to at least the policy level of the lock.}

Consider again our example from Section~\ref{sec:language}.
Recall that we had a list of papers \mbox{\AcceptedPapers} containing papers which were protected by policy $\Policy{\Clause{\set{\Published}}{\Pub}}$ and a location \Out{} storing data with policy~$\Policy{\Clause{\EmptyLock}{\Pub}}$.
For simplicity, we give papers the type $\Nat$.
Our original (insecure) implementation was the following:
\begin{align*}
	\Publish &\triangleq \Lam{x}{\Assgn{\Out}{x}}\\
	\PublishAll &\triangleq \App{\App{\LApply}{\Publish}}{\AcceptedPapers}
\end{align*}
If we type \textsf{publish} as $\ArrowType{\set{\Published}}{\set{\Clause{\EmptyLock}{\Pub}}}{\AnnotateType{\Nat}{\set{\Clause{\set{\Published}}{\Pub}}}}{\AnnotateType{\Unit}{\PolicyBot}}$, then it may only be invoked where the lock \Published is open.
Because this is not the case above, the type system rejects this implementation\added{ of \mbox{\PublishAll}}.
In order for the implementation of \PublishAll to be accepted by the type system we can change it as follows:
$$
\ExprIf*{\DateVar > \PubDate}
{\Open*{\Published}{\App{\App{\LApply}{\Publish}}{\AcceptedPapers}}}
{()}
$$
Alternatively, we can check \Published in \textsf{publish}, leaving the client \textsf{publishAll} as is.
In this case, \Publish would be defined as follows:
$$
\TypingPair{
  \Lam{x}
  {
    \ExprIf*{\DateVar > \PubDate}
    {\Open*{\Published}{\Assgn{\Out}{x}}}
    {()}
  }
}
{
  \ArrowType{\EmptyLock}{\set{\Clause{\EmptyLock}{\Pub}}}
  {\AnnotateType{\Nat}
    {\Policy{\Clause{\set{\Published}}{\Pub}}}}{\AnnotateType{\Unit}{\PolicyBot}}
}$$
As we can see from the type, this function can be invoked without any open locks because it opens the relevant lock itself.
Note that the type system does not prevent the programmer from opening \Published without checking that the publication date has passed, but by forcing the programmer to add code to open the lock, our type system prevents accidental declassification and forces the programmer to think about allowed information flows.
\added{Finally, we can also use the \mbox{\WhenKW} construct to define \mbox{\Publish} as follows:}
$$
\addedmaths{\DefMacro{\Publish}
	{
		\Lam*{x}
		{
			\When*{\Published}
			{\Assgn{\Out}{x}}
			{()}
		}
}}
$$
\added{In this case, \mbox{\Publish} also has type $\ArrowType{\EmptyLock}{\set{\Clause{\EmptyLock}{\Pub}}}
	{\AnnotateType{\Nat}
		{\Policy{\Clause{\set{\Published}}{\Pub}}}}{\AnnotateType{\Unit}{\PolicyBot}}$ and can be invoked in a context without the \mbox{\Published} lock open.}

% !TeX spellcheck = en_US
\section{A compositional definition of security}
\label{sec:comp-defin-secur}

We now turn to the main contribution of this paper: a compositional definition of security for \calcname{}.
Intuitively, we want our security definition to say that programs provide noninterference until a \emph{relevant declassification}, i.e., an allowed declassification that gives new information to an attacker.
More precisely, if we have two low-equivalent runs --- i.e., executions the attacker cannot tell apart --- then every reduction step in either execution that is not a relevant declassification should preserve that low equivalence.
However, once a relevant declassification has taken place the runs can diverge arbitrarily as that revealed information percolates through the execution.
For example in the program $\App{(\Lam{x}{\ExprIf{\Deref{l}}{\App{f}{\UnitVal}}{\App{g}{\UnitVal}}})}{(\Open{\LockSigmai{h}}{\Assgn{l}{\Deref{h}}})}$,
which first declassifies the value stored in $h$ to $l$ and then branches on that value, either $f$ or $g$ gets executed depending on the declassified value.

In effect, the description above captures a step-by-step version of the requirements posed by  Flow-Lock security~(\cite{BrobergS09}), from which we take inspiration.
In order to build a compositional definition of security that matches this intuition we use \emph{logical relations}, a standard tool in the semantics of programming languages.
We are not the first to use logical relations to define security compositionally.
Indeed, our semantics is built on that of FG~(\cite{RajaniG18}), which defines noninterference (without declassification) compositionally using logical relations.
However, as far as we are aware ours is the first model of where declassification using logical relations.
In fact, we believe this to be the first model of an intensional information flow property using logical~relations.

Logical relations provide models for programming languages by associating each type with a relation on terms.
Like most logical relations for information-flow security, our primary logical relation is \emph{binary}, meaning that its constituting relations each relate pairs of terms.
Intuitively, two terms are related if they are indistinguishable to a particular attacker~$\AttackerA$ \emph{until a relevant declassification occurs}.
We also use a \emph{unary} relation which tells us that programs do not perform low-visible effects in a high context.
In other words, the unary relation is the semantic analogue of a \emph{confinement} lemma.
Since our unary relation is adapted nearly directly from FG---though changed to a small-step relation---we postpone the description of it to Section~\ref{sec:unary-relation}, where we will only cover the~essentials.

Our binary relation---like most logical relations, including FG's---gives two interpretations of types: a \emph{value} relation and an \emph{expression} relation.
The value relation is written $\vb{\TypeTau}$, where $\AttackerA$ is an attacker (and $\mathcal{V}$ stands for ``value'').
Intuitively, the value relation tells us when two values are indistinguishable to an attacker.
Importantly, if the values are functions, we check that calling those functions results in indistinguishable programs.
This check provides the compositionality we set out to achieve.

The expression relation is written $\eb{\TypeTau}$.
Again, $\AttackerA$ is an attacker while $E$ stands for ``expression.''
Intuitively, the expression relation checks that two expressions evaluate to indistinguishable values and that information is never leaked to an adversary during the evaluation.
If evaluating the expressions results in a relevant declassification, i.e., a pair of declassification steps that leads to diverging behavior, we no longer make any requirements on the expression.
Importantly, this does not mean that our security definition does not detect information leaks beyond a fixed declassification point in the program. Our security definition considers \emph{all pairs} of executions. Given any declassification point in a program, even if one pair of executions results in a relevant declassification because the two executions declassify different data at that point, in other pairs of executions the two executions may declassify the same data at that point making the declassification non-relevant in those pairs of executions, and the security definition's requirements would continue to be imposed past that point for those pairs.

\mnb{Attackers}
In order to reason about what an attacker can see, we need to define attackers.
Usually in information-flow--security definitions, an attacker is an information-flow label.
However, \calcname{} uses Flow-Locks policies, which may contain more than one actor, rather than information-flow labels.
In order to maintain the intuition of a single attacker, we follow Flow Locks~(\cite{BrobergS09,BrobergS10}) in defining an attacker~$\AttackerA$ as a policy $\set{\Clause{\AttackerLocks}{\A}}$.
That is, an attacker~$\AttackerA$ is a pair $(\A,\AttackerLocks)$, of an actor~$\A$ and a lock set~$\AttackerLocks$ called the \emph{capability} of~$\AttackerA$.

We can intuitively think of $\AttackerLocks$ as the set of locks that the attacker can open forcibly.
Accordingly, a policy $\PolicyP$ is visible to an attacker when $\PolicyOrder{\PolicyP}{\set{\Clause{\AttackerLocks}{\A}}}$ (abbreviated $\PolicyOrder{\PolicyP}{\AttackerA}$).
This occurs if $\A$ may learn information protected by $\PolicyP$ under the assumption that the locks in $\AttackerLocks$ are open.

\mnb{State indistinguishability}
We want two expressions to be indistinguishable to an attacker~$\AttackerA$ if, when started in $\AttackerA$-indistinguishable states, the two traces of observations produced by their execution and the final reduced values are also $\AttackerA$-indistinguishable.
This requires us to determine when two states are $\AttackerA$-indistinguishable.

If \calcname{} had no way to allocate new memory cells, then we could simply say that two states are indistinguishable if every state cell contained indistinguishable values.
More precisely, if $\theta$ was a state environment, and $S_1, S_2$ were two states of type~$\theta$, then $S_1$ and $S_2$ would be $\AttackerA$-indistinguishable if \mbox{$(S_1(l), S_2(l)) \in \vb{\theta(l)}$} for every location~$l$.

However, since \calcname{} has nondeterministic memory allocation, things get more complicated.
For example, suppose we wish to prove that $e_1 \triangleq \Let{x}{\New(1,\AnnotateType{\Nat}{\bot})}{\Deref{x}}$ is related to itself.
The two copies of $e_1$ may allocate different locations, say $l_1$ and $l_2$, at the subexpression $\New(1,\AnnotateType{\Nat}{\bot})$.
Consequently, one reduction step later, we would want to relate $\Deref{l_1}$ and $\Deref{l_2}$ in the starting states $l_1 \mapsto 1$ and $l_2 \mapsto 1$.
These two states do not have the same locations, yet, somehow, we wish to say that they are indistinguishable.

In order to handle this, we adopt a standard solution: we allow each run to have different state environments, as long as there is a bijection between corresponding locations in the two runs.
More precisely, we index all of our relations with Kripke \emph{worlds}~(\cite{AhmedDR09}) (usually denoted by W, W', etc.), which are triples of the form $(\theta_1, \theta_2, \beta)$ where $\theta_1$ and $\theta_2$ are state environments and $\beta$ is a bijection between their domains.
The bijection $\beta$ tells us how to match up locations between the runs.
In our example above, when we get to relating~$\Deref{l_1}$ and~$\Deref{l_2}$, the bijection $\beta$ would match~$l_1$ and~$l_2$, while $\theta_1$ and $\theta_2$ would assign~$l_1$ and~$l_2$, respectively, $\AnnotateType{\Nat}{\bot}$. 
Clearly, the expressions $\Deref{l_1}$ and $\Deref{l_2}$ will both evaluate to the same value in any two states that are indistinguishable up to this $\beta$, which justifies their relatedness.

So far, this definition implicitly assumes that when the program on the left-hand side of the relation allocates memory, so will the program on the right-hand side.
After all, a bijection between two sets witnesses that they have the same size.
This assumption does not hold, however, since secrets may determine whether a location is allocated.
To see this, consider proving that $e_2 \triangleq \ExprIf{\Deref{h}}{\Let{x}{\New(3,\AnnotateType{\Nat}{\top})}{\Deref{x}}}{3}$ is related to itself, where $h$ is a high-security reference.
Since $h$ is high security, in one run it may store true while in the other run it stores false.
Thus, after two steps of computation, we need to relate $\Let{x}{\New(3,\AnnotateType{\Nat}{\top})}{\Deref{x}}$ with $3$.
On the left-hand side we allocate memory that we do not on the right.
However, since this is a private memory cell, intuitively the adversary cannot see that allocation.

In order to accommodate situations like the one above, where differing numbers of private allocations occur in two runs, we relax the requirement that $\beta$ be a bijection.
Instead, we only require that it be a \emph{partial} bijection; that is, an injective partial function.
We consider those locations matched by $\beta$ to be ``the same,'' while those not related to any location by $\beta$ correspond to situations like that in $e_2$.
Moreover, our binary relation requires that attacker-visible locations allocated in both branches be matched by $\beta$.
Combining these facts, only private locations will fail to be related to any other location.

Formally, the $\AttackerA$-indistinguishability of states $S_1$ and $S_2$ at world $W$ is denoted $(S_1,S_2,\unimp{m}) \bwf (W)$, and is defined in Definition~\ref{def:state-indist}.
The part in \unimptext{gray}, called the step index (\cite{AppelM01}), can be ignored by most readers.
We discuss it later.
\begin{definition}[World-indexed state indistinguishability]~\\
  \label{def:state-indist}
  We say that states~$S_1$ and~$S_2$ are \emph{$\AttackerA$-indistinguishable in world~$W$}, written $(S_1,S_2, \unimp{m}) \bwf W$, exactly when:
  \begin{itemize}
  \item[] $\Dom(W.\theta_i) \subseteq \Dom(S_i)$
  \item[$\mathrel{\land}$] $\forall l \in
    \Dom(W.\theta_i). \, W.\theta_i(l) = \HeapTypeLookup(S_i,l)$
  \item[$\mathrel{\land}$] $\beta \subseteq \Dom(W.\theta_1) \times \Dom(W.\theta_2)$
  \item[$\mathrel{\land}$]  $\left(
      \begin{array}[c]{l}
        \forall (l_1, l_2) \in W.\beta.\\
        ~~W.\theta_1(l_1) = W.\theta_2(l_2) \mathrel{\land} \\
        ~~(S_1(l_1),S_2(l_2),W,\unimp{m}) \in \vb{W.\theta_1(l_1)}
      \end{array}\right)$
  \end{itemize}
\end{definition}

The most important part of this definition is its last line, which says that for all locations $l_1$ and $l_2$ matched by $\beta$, the values in those locations must be indistinguishable at their common type.
Again, we represent indistinguishability of values using the value interpretation $\vb{W.\theta_1(l_1)}$, but the astute reader will note that this interpretation seems to relate the \emph{quadruple} $(S_1(l),S_2(l),W,\unimp{m})$ rather than the pair $(S_1(l),S_2(l))$.
This is because the definition of our value interpretation $\vb{W.\theta_1(l_1)}$ also includes worlds.
These worlds are necessary because function values may contain free locations, which could differ in the two runs.
Roughly, \mbox{$(v_1, v_2, W,\unimp{m}) \in \vb{\TypeTau}$} means that $v_1$ and $v_2$ are $\AttackerA$-indistinguishable, up to renaming of locations under~$W.\beta$.
The full definition of $\vb{\TypeTau}$ can be found in Section~\ref{sec:value-relation}.
Our notions of worlds and state indistinguishability presented above correspond to the ones in FG~(\cite{RajaniG18}).

\subsection{The expression relation}
\label{sec:expression-relation}
We now begin to explain the formal definition of indistinguishability through our logical relation.
The complete definition of the expression relation can be found in Figure~\ref{fig:binary-exp} (on the next page), while the value relation can be found in Figure~\ref{fig:binary-value-rel} (in Section~\ref{sec:value-relation}).

In order to explain the expression relation, which looks quite complicated, we break the explanation up into three parts.
In the first part, we explain the parts of the relation relevant to noninterference.
This part of the relation acts similarly to the relation for FG.
However, in order to accommodate declassification, we have changed the presentation of the relation to use small-step operational semantics, rather than big-step.
Then, we describe the parts of our relation that are relevant to declassification.
Finally, we touch briefly on our use of step indexing (\cite{AppelM01}), which is a technique for ensuring that logical relations are well defined in the face of higher-order state.
In order to make the following sections easier to read, we have highlighted the parts of the relation related to declassification \newtext{in yellow} and those related to step indexing \unimptext{in gray}.
Thus, the first part of this subsection focuses on the \emph{unhighlighted} parts of the expression relation.
Then, we focus on the parts \newtext{in yellow}.
Finally, we touch briefly on the parts \unimptext{in gray}.

\begin{figure*}[t]
	\tiny
	\providecommandx{\lred}[7][2=W',3=\LockSetSigmaI,4=m',5=\LockSetSigma,6=\LockSetSigmaPrime,7=\LockSetSigmaPrimeI,usedefault]{\comprehend{(e_1,e_2)}{
			\begin{array}{l}
				e_1 \notin \mathcal{V}\ \land  \forall e'_1, S'_1, \newstuff{#7}, \ObservationOmega. \\[0.75em] \Step{e_1,\newstuff{#3},S_1, e'_1,S'_1, \smash{\ObservationOmega},\smash{\newstuff{#7}}} \rightarrow
				\NotPolicyOrder{\ObsPolicy(\ObservationOmega)}{\AttackerA} \ \land \\
				(\exists #2'.#2' \sqsupseteq #2 \land (S'_1,S_2,\unimp{#4}) \overset{\AttackerA}{\triangleright} (#2') \ \land\\  (e'_1,e_2,#2',\newstuff{#5},\newstuff{#6},\unimp{#4}) \in \eb{#1})
			\end{array}
	}}
	
	\providecommandx{\rred}[7][2=W',3=\LockSetSigmaII,4=m',5=\LockSetSigma,6=\LockSetSigmaPrime,7=\LockSetSigmaPrimeII,usedefault]{\comprehend{(e_1,e_2)}{
			\begin{array}{l}
				e_2 \notin \mathcal{V}\ \land \forall e'_2, S'_2, \newstuff{#7}, \ObservationOmega. \\[0.75em]
				\Step{e_2,\newstuff{#3},S_2, e'_2,S'_2,\smash{\ObservationOmega},\smash{\newstuff{#7}}} \rightarrow \NotPolicyOrder{\ObsPolicy(\ObservationOmega)}{\AttackerA} \ \land \\
				(\exists #2'. #2' \sqsupseteq #2 \land (S_1,S'_2,\unimp{#4}) \bwf #2' \ \land\\  (e_1,e'_2,#2',\newstuff{#5},\newstuff{#6},\unimp{#4}) \in \eb{#1})
			\end{array}
	}}
	
	\begin{align*}
		&\eb{\TypeTau} \triangleq \ebb{\TypeTau} \cup \set{(v,v',W,\newstuff{\LockSetSigmai{c_1}},\newstuff{\LockSetSigmai{c_2}},\unimp{m}) \where (v,v',W,\unimp{m}) \in \vb{\TypeTau}} \mbox{ where:} \\\\
		&\ebb{\TypeTau} \triangleq 
		\comprehend{(e_1,e_2,W,\newstuff{\LockSetSigma},\newstuff{\LockSetSigmaPrime},\unimp{m})}
		{
			\begin{array}{l}
				\newstuff{\addedmaths{\LockSetSigma \lowequiv \LockSetSigmaPrime \land\ } \forall \LockSetSigmaI, \LockSetSigmaII. \LockSetSigma \subseteq \LockSetSigmaI \land \LockSetSigmaPrime \subseteq \LockSetSigmaII \addedmaths{\land\ \LockSetSigmaI \lowequiv \LockSetSigmaII} \rightarrow}
				\forall W',\unimp{m'}, S_1,S_2.\\ \unimp{m' < m} \land W' \sqsupseteq W \land (S_1,S_2,\unimp{m'}) \bwf W' \rightarrow (e_1,e_2) \in \\~\\
				\left(
				\begin{array}{l}
					C_{\textsc{Par}} = \comprehend{(e_1, e_2)}{
						\begin{array}{l}
							e_1 \notin \mathcal{V}\ \land\ %
							e_2 \notin \mathcal{V}\ \land \\ %
							\forall e'_1, S'_1, \newstuff{\LockSetSigmaPrimeI},\ObservationOmega,e'_2, \newstuff{\LockSetSigmaPrimeII}, S'_2, \ObservationOmegaPrime. \\[0.5em]
							\Step{e_1,\newstuff{\LockSetSigmaI},S_1,e'_1,S'_1,\smash{\ObservationOmega},\newstuff{\LockSetSigmaPrimeI}}\ \land \\
							\Step{e_2,\newstuff{\LockSetSigmaII},S_2,e'_2,S'_2, \smash{\ObservationOmegaPrime},\smash{\newstuff{\LockSetSigmaPrimeII}}   }\rightarrow\\
							\newstuff{%
								\addedmaths{(\ObservationOmega \equivWorld[W',m'] \ObservationOmegaPrime \lor \PolicyOrder{\LockSetSigmaPrimeI}{\AttackerA}  \lor  \PolicyOrder{\LockSetSigmaPrimeII}{\AttackerA} )
								\rightarrow}
								}\\
							\exists W''. W'' \sqsupseteq W' \land\ (S'_1,S'_2,\unimp{m'}) \bwf (W'') \ \land \\  \ObservationOmega \equivWelt[(W'',\unimp{m'})] \ObservationOmegaPrime \land
							\\ (e'_1,e'_2,W'',\newstuff{\LockSetSigma},\newstuff{\LockSetSigma'},\unimp{m'}) \in \eb{\TypeTau}
					\end{array}} \cup\\ ~\\
					C_{\textsc{L}} =\lred{\TypeTau}
					\cup\\ ~\\
					C_{\textsc{R}} = \rred{\TypeTau}
				\end{array}\right)
			\end{array}
		}\\
	\end{align*}
	\caption{Binary expression relation}
	\label{fig:binary-exp}
\end{figure*}

\mnb{Relating expressions without declassification}
The expression relation $\eb{\cdot}$ is a union of two relations: a relation $\ebb{\cdot}$ (from here on \added{reduction }\del{$\beta$-}relation in the text) when at least one of the expressions can take a step; and the value relation, when both of the expressions are values.
Thus, the logical relation enforces progress: stuck programs are never in the expression relation.
We describe the value relation in Section~\ref{sec:value-relation}; the rest of this subsection will focus on the \added{reduction}\del{$\beta$} relation.

In the \added{reduction}\del{$\beta$} relation, we start by picking an arbitrary \emph{future world} $W' \sqsupseteq W$ and two arbitrary states $S_1$ and $S_2$ which are $\AttackerA$-indistinguishable in $W'$.
Intuitively, the choice of future world says that the relation must continue to hold even as the state evolves.
This is an important ingredient of compositionality: functions that are considered equivalent when they are defined should continue to be equivalent when they are used, even if more locations have been allocated.
By ensuring that our relation holds in an arbitrary future world, we make our relation \emph{monotonic}: if $e_1$ and $e_2$ are related at some world~$W$, then $e_1$ and $e_2$ remain related in any future~world~$W' \sqsupseteq W$.
In the rest of this subsection\added{,} we will see that ensuring such monotonicity properties is an important goal throughout the definition of the \added{reduction}\del{$\beta$} relation, and in the definition of the logical relation as a whole.
\begin{definition}[Future World~(\cite{RajaniG18})]
  Formally, we say \emph{$W'$ is a future world of $W$} (written $W' \sqsupseteq W$) whenever \mbox{$W.\theta_i \subseteq W'.\theta_i$} for $i = 1, 2$ and $W.\beta \subseteq W'.\beta$.
  Thus, all allocated locations remain allocated with the same type, and locations which are matched across the two runs continue to be matched.
\end{definition}

Intuitively, our relation says that the attacker~$\AttackerA$ cannot distinguish the executions of $e_1$ and $e_2$.
We thus require that those execution steps of $e_1$ and $e_2$ that the attacker can observe be matched one-to-one.
However, we allow $e_1$ and $e_2$ to take any number of invisible steps with no match on the other side.
In order to formalize this intuition, we split the \added{reduction}\del{$\beta$} relation into three parts: $C_{\textsc{Par}}$, $C_{\textsc{L}}$, and $C_{\textsc{R}}$.
The $C_{\textsc{Par}}$ relation corresponds to the requirement that observable steps be matched one-to-one.
The $C_{\textsc{L}}$ and $C_{\textsc{R}}$ relations correspond to $e_1$ and $e_2$, respectively, taking unmatched invisible steps.

The $C_{\textsc{Par}}$ subrelation requires that neither $e_1$ nor $e_2$ be values, since values cannot take any steps.
Then, any steps of $e_1$ and $e_2$ result in new expressions~$e_1'$ and~$e_2'$ and new states~$S_1'$ and $S_2'$.
In order for $e_1$ and $e_2$ to be indistinguishable, the adversary should not be able to distinguish $S_1'$ and $S_2'$.
However, the step we just took may have allocated more locations, so we cannot require that $S_1'$ and $S_2'$ be indistinguishable in $W'$.
Thus, we require that there exist some future world~$W'' \sqsupseteq W'$ such that $S_1'$ and $S_2'$ are indistinguishable.

\begin{figure}
    \begin{mathpar}
      \Infer{high}{\NotPolicyOrder{\ObsPolicy(\ObservationOmega)}{\AttackerA} \\ \NotPolicyOrder{\ObsPolicy(\ObservationOmegaPrime)}{\AttackerA}}{\ObservationOmega \equivWelt[(W,\unimp{m})] \ObservationOmegaPrime} \and
            \addedmaths{
      	\Infer{refl}{\forall l, \TypeTau, v.\ \ObservationOmega \neq \WriteObs{l}{\TypeTau}(v)}{
      		\ObservationOmega \equivWelt \ObservationOmega}
      	} \and
\Infer{extend-$\tau$}{(v,v',W,\unimp{m}) \in \vb{\TypeTau} \\ (l,l') \in W.\beta}{\WriteObs{l}{\TypeTau}(v) \equivWelt[(W,\unimp{m})] \WriteObs{l'}{\TypeTau}(v')}

    \end{mathpar}
  \caption{Observational indistinguishability}
  \label{fig:obs-indist}
\end{figure}

Next we require that the observations $\ObservationOmega$ and $\ObservationOmegaPrime$ generated by the steps above be indistinguishable to the attacker~$\AttackerA$.
Most security definitions just require that observations are equal.
Because we are in the higher-order setting and wish to get a compositional definition, we need to be a bit more careful.
We formally define observation indistinguishability in Figure~\ref{fig:obs-indist}.
Intuitively, two observations are indistinguishable if the attacker cannot see either of them, as the rule~\textsc{high} expresses, or if they look the same to the attacker.
\added{In most cases, observations look the same to an attacker if they are in fact the same,
  as codified by the rule~\textsc{refl}.
The exception to this rule are write observations.}
If \del{the attacker can see them,} both observations are writes---i.e., we are trying to prove $\WriteObs{l}{\TypeTau}(v) \equivWelt[(W,\unimp{m})] \WriteObs{l}{\TypeTau}(v')$\del{. In this case~}\added{, }\textsc{extend-$\tau$} applies.
To ensure that both writes look the same to the attacker, the locations~$l$ and~$l'$ must be related by $W.\beta$, since they must represent ``the same'' location between the two runs.
Additionally, $v$ and $v'$ must be indistinguishable to $\AttackerA$.

Returning to the definition of $C_{\textsc{Par}}$ in Figure~\ref{fig:binary-exp}, we finally require that the new programs $e_1'$ and $e_2'$ are indistinguishable in this new world $W''$.
Thus, any further steps they take are again subject to the requirements of the \added{reduction}\del{$\beta$} relation, ensuring that security holds for more than one step.

We now look at the left-hand relation $C_{\textsc{L}}$; the right-hand relation $C_{\textsc{R}}$ is the dual, so we avoid explicitly describing it.
Again, $e_1$ should not be a value, since values are not allowed to take steps.
However, $e_2$ may now be a value, since we do not consider any steps it may take.
In order to be in the left-hand relation, we require that $e_1$ only be able to take steps which are \emph{not} visible to the attacker~$\AttackerA$.
In other words, every step that $e_1$ can take should yield an observation $\ObservationOmega$ such that $\NotPolicyOrder{\ObsPolicy(\ObservationOmega)}{\AttackerA}$.
(Note that $\EmptyObs$ is one such observation.)
When $e_1$ takes such a step, yielding a new expression~$e_1'$ and a new state~$S_1'$, intuitively the attacker should not be able to tell that $e_1$ took a step.
To formalize this, we require that $S_1'$ is indistinguishable from $S_2$ at some future world $W'' \sqsupseteq W'$.
We also require that $e_1'$ be indistinguishable from $e_2$, again ensuring that security holds for more than one step.

Note that our relation enforces a \emph{termination insensitive} indistinguishability relation:
we consider a program $e_t$ that terminates to be indistinguishable from another program $e'$ that continues executing after $e_t$'s termination but has been indistinguishable from $e_t$ up until this point.
We choose a termination insensitive definition because our type system only enforces termination insensitive security.
Termination-sensitive type systems are notoriously restrictive and difficult to work with.
We could turn our logical relation into a termination-sensitive definition by enforcing in $C_{\textsc{Par}}$ that any step in the left execution is matched by a step in the right execution and by enforcing that $C_{\textsc{L}}$ and $C_{\textsc{R}}$ can only be used a finite number of times before another clause is used.

\mnb{Handling where declassification}
We now revisit the definition of the expression relation, considering the parts \newtext{in yellow} which correspond to reasoning about declassification.
Most of the changes involve carrying around two sets of open locks~$\LockSetSigmaI$ and~$\LockSetSigmaII$, one for each expression.
Notably, the value relation does not require us to reason about open locks, since the only way values can contain declassifications is inside functions, which carry their open-lock sets with them.
Thus, values are related at all pairs of lock sets if they are related at all.
With this insight, we can again focus our attention on the $\beta$~relation.

We begin by \added{enforcing that the two lock sets are indistinguishable to the attacker, i.e., the locks visible to an attacker are identical.
Formally we define the set of locks in $\LockSetSigma$ visible to an attacker $\AttackerA$ as $\low{\LockSetSigma} := \set{\LockSigma \in \LockSetSigma \where \LockPolicy(\LockSigma) \sqsubseteq \AttackerA}$.
We say that two lock sets $\LockSetSigma$ and $\LockSetSigmaPrime$ are low equivalent for attacker $\AttackerA$ (written $\LockSetSigma \lowequiv \LockSetSigmaPrime$) if $\low{\LockSetSigma} = \low{\LockSetSigmaPrime}$.
By ensuring we have low-equivalent lock sets we ensure that any difference observed by the attacker is actually due to secret information being released and not just because of an initial difference in the opened locks.
We then continue by }picking \added{low-equivalent} supersets~$\LockSetSigmaI$ and~$\LockSetSigmaII$ of our open lock sets $\LockSetSigma$ and~$\LockSetSigmaPrime$.
This corresponds to the intuition that a term that is deemed secure starting from a certain set of open locks should also be deemed secure starting from a larger set of open locks (since having more open locks can only make more, not fewer, declassifications policy compliant).
Again, this leads to a \emph{monotonicity} property: if two terms are related with lock sets~$\LockSetSigma$ and~$\LockSetSigmaPrime$, then they are related with any supersets of those lock~sets.

Once we consider lock sets, we must index every step of evaluation with the current open locks and with the active lock set of the reduction.
In $C_{\textsc{L}}$ and $C_{\textsc{R}}$, these play no further role in the logical relation; since lock sets are static, the original lock sets are reused in recursive calls.
However, in $C_{\textsc{Par}}$ there is an additional clause which needs explanation.

The key difference between noninterference and declassification is that two executions of noninterfering programs are \emph{always} indistinguishable, while \textbf{executions of programs with declassification may stop being indistinguishable at some point.}
To see this, consider two executions of a program that differ only in secret inputs.
Once a secret is declassified, the two executions now differ in both public and secret data.
Thus, they may no longer behave the same and may return distinguishable outputs.
As long as the declassification that caused the executions to get out of sync was allowed according to policy, this is not a security breach.

We call a declassification that can cause the two executions to legally get out of sync a \emph{relevant declassification}.
Because declassification happens during the execution of the program, rather than as part of a value, the formalization of relevant declassification appears as part of the expression relation.
Once a relevant declassification appears, we need to stop requiring that the programs act in lockstep.
It is precisely to formalize this that we have based our logical relation on small-step reduction, considering each step in isolation in order to establish a lockstep bisimulation.
When formalizing an extensional property like noninterference, this is unnecessary; thus, the structure of our expression relation is significantly different from FG's logical relation.

We formalize relevant declassifications as part of the $C_{\textsc{Par}}$~clause in the expression relation.
\added{To explain relevant declassification and its use in the $C_{\textsc{Par}}$~clause, we first consider a different but equivalent formulation of the $C_{\textsc{Par}}$~clause from that one shown in Figure~\mbox{\ref{fig:binary-exp}}.}
\[\addedmaths{\comprehend{(e_1, e_2)}{
		\begin{array}{l}
			e_1 \notin \mathcal{V}\ \land\ %
			e_2 \notin \mathcal{V}\ \land \\ %
			\forall e'_1, S'_1, \newstuff{\LockSetSigmaPrimeI},\ObservationOmega,e'_2, \newstuff{\LockSetSigmaPrimeII}, S'_2, \ObservationOmegaPrime. \\[0.5em]
			\Step{e_1,\newstuff{\LockSetSigmaI},S_1,e'_1,S'_1,\smash{\ObservationOmega},\newstuff{\LockSetSigmaPrimeI}}\ \land \\
			\Step{e_2,\newstuff{\LockSetSigmaII},S_2,e'_2,S'_2, \smash{\ObservationOmegaPrime},\smash{\newstuff{\LockSetSigmaPrimeII}}   }\rightarrow\\
			\newstuff{%
				\left[
				\begin{array}{l}
					\addedmaths{\ObservationOmega \neq \EmptyObs \land \ObservationOmegaPrime \neq \EmptyObs \land} \\
					\addedmaths{(\ObservationOmega \neq \ObservationOmegaPrime \land \forall l, v, \TauPrime.\ (\ObservationOmega \neq \WriteObs{l}{\TauPrime}(v) \lor\ObservationOmegaPrime \neq \WriteObs{l}{\TauPrime}(v)) \lor}\\
					\exists l,v,\TauPrime,l',v'. \\
					\ObservationOmega = \WriteObs{l}{\TauPrime}(v) \land \ObservationOmegaPrime = \WriteObs{l'}{\TauPrime}(v')\ \land\\
					(v, v', W', \unimp{m'}) \notin \vb{\TauPrime}\ \addedmaths{)}\land\\
					\PolicyOrder{\ObsPolicy(\ObservationOmega)}{\AttackerA} \land \PolicyOrder{\ObsPolicy(\ObservationOmegaPrime)}{\AttackerA}\ \land\\
					\NotPolicyOrder{\LockSetSigmaPrimeI}{\AttackerA} \land \NotPolicyOrder{\LockSetSigmaPrimeII}{\AttackerA}
				\end{array}
				\right] \lor}\\
			\exists W''. W'' \sqsupseteq W' \land\ (S'_1,S'_2,\unimp{m'}) \bwf (W'') \ \land \\  \ObservationOmega \equivWelt[(W'',\unimp{m'})] \ObservationOmegaPrime \land
			\\ (e'_1,e'_2,W'',\newstuff{\LockSetSigma},\newstuff{\LockSetSigma'},\unimp{m'}) \in \eb{\TypeTau}
\end{array}}}
\]

\added{
  The main difference between this and the version of $C_{\textsc{Par}}$ without declassification (other than the use of lock sets) is the large square-bracketed block $[\ldots]$.
  This block is the formalization of relevant declassification.
  At a high level, the above $C_{\textsc{Par}}$ clause says that if we have a pair of reductions, we can either establish that the produced observations and the resulting states and programs are indistinguishable, as we would do for noninterference, or alternatively, we can show that these two executions constitute relevant declassification.}

There are four major requirements in order for a step to be a relevant declassification, corresponding to the four lines of the \newtext{large yellow \del{conjunction}\added{block}} in the definition of $C_{\textsc{Par}}$.
First, the two steps must \added{be}\del{write to memory, providing} something that \del{the}\added{an} adversary could possibly see\added{---in particular, they cannot be $\EmptyObs$}.
\del{We enforce this by requiring that the observations must be write observations.
}

Second, those \del{writes}\added{observations} must be \del{to a location that}\added{visible to} the \added{specific }attacker \added{under consideration}\del{can see}; we formalize this by requiring that the policy of the observation flow to the attacker.
Not only does this make intuitive sense (if the attacker cannot see a\added{n} \del{location}\added{observation}, \del{writing data to that location}\added{it} does not release any information to the attacker), but it is also vital for enforcing security.
To see this, consider the program $(\Open{\LockSigmai{h}}{\Assgn{m}{h}}); \Assgn{l}{h}$ where $l$ is not protected by locks, while $m$ and $h$ are protected by locks~$\LockSigmai{m}$ and~$\LockSigmai{h}$, respectively.
This program first declassifies $h$ to $m$, and then \emph{leaks} $h$ to $l$.
Since any attacker which can see neither $m$ nor $h$ sees new information when $h$ is assigned to $l$, our logical relation correctly rules out this program.
If we considered the declassification of $m$ to $h$ to be relevant, however, we would stop requiring that the programs be indistinguishable after $\Assgn{m}{h}$ and thus incorrectly deem this program secure.

Third, the attacker must be able to distinguish the observations in the two executions.
\added{Because write observations are more complicated than the other observations, we distinguish two cases:
either at least one observation is not a write observation or both observations are write observations.
In the first case, it suffices for the observations to be different.
In the second case we need to ensure the values we write in both executions are distinguishable.}
Because our notion of \emph{in}distinguishability is being related in the logical relation, we formalize distinguishability by requiring that the two values being written are \emph{not} related by the logical relation.

It may be surprising that we do not consider observations that write the same value to different locations (i.e., locations that are not related by the partial bijection $\beta$) distinguishable.
This has the consequence that our definition deems certain programs insecure that we would consider secure if we did take the locations into account.
For example if the location $h$ is protected by a lock $\LockSigmai{h}$ and $l, l'$ are publicly visible, then the program $\ExprIf{\Deref{h}}{(\Open{\LockSigmai{h}}{\Assgn{l}{5}})}{(\Open{\LockSigmai{h}}{\Assgn{l'}{5}})}$ is deemed insecure by our definition since the writes of $5$ to $l$ and $l'$ are not deemed relevant declassification but the program still leaks the value of $h$.

The obvious question is why our definition of relevant declassification chooses to ignore the relatedness of locations that were written.
The reason is there is no world whose $\beta$ we can use to check the relatedness of the locations correctly across all programs in this situation.
If we were to use the world prior to the parallel steps (i.e., $W'$), a pair of locations that is newly allocated during the step would not be related under this world's $\beta$ when they should be.
On the other hand, we cannot yet extend to a future world, since there will always be a future world in which the newly allocated locations are not related, so any pair of steps allocating new locations would constitute relevant~declassification.

We could solve this problem by only taking into account locations already matched by $\beta$ and ignoring all other locations when checking relatedness, but we choose not to do so here to avoid complicating our definition further.
Note that our current choice is ``safe'' in the sense that it only deems fewer, not more, programs secure.
It also does not cause any trouble in proving our type system sound since this situation can only occur if
the program has previously branched on a secret value and the type system, like most other information flow type systems, rules out public writes in branches influenced by non-public values (as in the example above). In particular, the example would not type in our -- and most existing -- type systems.

Fourth, the currently open locks must \emph{allow} declassification, otherwise a difference in observed behavior would just be a leak. We ensure that declassification is allowed by checking that the open locks at the point of the declassification are not all contained in the attacker's capability. Indeed, if the open locks were contained in the attacker's capability, then the attacker should not learn anything new from the declassification so the declassification would not be relevant.
On first glance just requiring that some lock is open that is not part of the attacker's capability might seem too weak.
Intuitively one might expect that we require that the correct locks needed to declassify the value in question, namely all locks in the \emph{difference} of the lock set that must be open for the attacker to see the value that the step writes and the lock set that must be open for the attacker to see the location that is written, are actually open.
Without this condition, the declassification is illegal and should not be considered relevant.
The reason it is sound to omit this condition here is that our program security definition (Definition~\ref{def:security}) requires security against \emph{all} attackers.
If a lock $\LockSigma$ required for the flow to be legal is not actually open, we can choose a different attacker $\AttackerA_{\LockSigma}$ whose capability excludes $\LockSigma$ but includes all locks needed to view the location (i.e., the observation of the step).
For this attacker, the step will not be relevant declassification and our security definition will deem the program insecure.

It might also be surprising that our relevant declassification condition occurs only in the $C_{\textsc{Par}}$ clause.
Indeed, one might expect that a program that only declassifies data in one execution also exhibits relevant declassification.
However, considering such a situation relevant declassification would incorrectly classify insecure programs as secure.
To see this, recall that we start executions in indistinguishable states.
Thus, if a declassification happens in one execution but not the other, that difference must come from branching on some secret value~$h$.
Therefore, if there was a declassification in one execution, the absence of a similar declassification in the other execution would leak information about $h$.

\del{Note that}\added{In the explanation above,} we use the formalization of relevant declassification in $C_{\textsc{Par}}$ as part of a disjunction.%IMPORTANT FOR SPACING; DO NOT REMOVE
\del{\footnote{\del{The technical~appendix actually uses an equivalent implication instead of a disjunction, as it is more convenient for formal reasoning.
	We chose to explain the equivalent formulation with a disjunction here for readability's sake.}
}}
In other words, every parallel step must show either that the step is a relevant declassification or that the bisimulation continues in lockstep.
This matches the key intuition for declassification that we gave earlier: declassification allows us to \emph{stop} requiring that two executions be bisimilar, while noninterference requires bisimilarity for the entirety of the traces.
\added{However, this is not how we actually formalize it in the logical relation.
Using a disjunction suggests that we have to decide for every pair of execution steps if they constitute relevant declassification.
This is not very practical, especially if we are reasoning about abstract programs---for example, when proving properties about the entire language.
Instead, we replace the disjunction expressing that \emph{there is either relevant declassification or continued noninterference} with the classically equivalent implication \emph{if there is no relevant declassification, there must be continued noninterference}.

Additionally, the $C_{\textsc{Par}}$ clause in the formal logical relation encapsulates the reasoning about observations in a separate proposition.
Note that relevant declassification encompasses two separate aspects.
First, it requires that declassification is allowed, which is captured by $\NotPolicyOrder{\LockSetSigmaPrimeI}{\AttackerA} \land \NotPolicyOrder{\LockSetSigmaPrimeII}{\AttackerA}$ as explained above.
Second, it ensures that the attacker has actually learned new information, which we formalize by requiring that the attacker was able to make two distinguishable observations.
In the implication formalization of the $C_{\textsc{Par}}$ clause, where we assume the absence of relevant declassification, this changes to assuming that the attacker has not learned new information, i.e., that they cannot tell the two observations apart.
It is, therefore, convenient to formalize this as an indistinguishability predicate ($\equivWorld$) on the two observations. This predicate is defined in Figure \mbox{\ref{fig:obs-irrel}}.
}
\begin{figure}
\begin{mathpar}
\addedmaths{	\Infer{\textbf{refl}}{
		\forall l,\TypeTau,v. \ObservationOmega \neq \WriteObs{l}{\TypeTau}(v)}{\ObservationOmega \equivWorld \ObservationOmega}} \and
	\addedmaths{\Infer{high}{\neg (\PolicyOrder{\ObsPolicy(\ObservationOmega)}{\AttackerA}) \lor \neg(\PolicyOrder{\ObsPolicy(\ObservationOmegaPrime)}{\AttackerA})}{\ObservationOmega \equivWorld \ObservationOmegaPrime}}
	\and
	\addedmaths{
	\Infer{extend-$\tau$}{(v,v',W,m) \in \vb{\TypeTau}}{\WriteObs{l}{\TypeTau}(v) \equivWorld \WriteObs{l'}{\TypeTau}(v')}}
\end{mathpar}
  \caption{\added{Observational irrelevance}}
\label{fig:obs-irrel}
\end{figure}
\added{As indicated by the rule names, this relation is very similar to observational indistinguishability (Figure \mbox{\ref{fig:obs-indist}}),
but it is slightly weaker, i.e., treats more observations as indistinguishable.
First, it ignores locations in write observations for the reasons explained above.
Second, it only requires one observation to be invisible for the \mbox{\rname{high}} rule to apply.
This ensures that invisible observations are never part of a relevant declassification.}

\mnb{Step indexing}
We now turn to the last part of our logical relation, the part of the relation \unimptext{in gray}.
Since our logical relation is defined recursively, in order for the relation to be well defined the recursion must be well founded.
Like FG~(\cite{RajaniG18}), we use a standard trick to ensure wellfoundedness, \emph{step indexing}~\cite[see~e.g.,][]{AppelM01,AhmedDR09,Ahmed03}.
Essentially, step indexing forces us to define a well founded recursion over the natural numbers.
By only making recursive calls at strictly smaller indices, we can be sure that we do not encounter any circularity.
This is similar to the trick commonly used when defining a terminating function: by supplying an extra (natural number) \emph{fuel} parameter to the function, we can ensure that the function terminates.

Compositionality again demands that we make the relation monotonic along decreasing $m$.
We do this in the \added{reduction}\del{$\beta$} relation by choosing an $m' < m$ when we want to show that $e_1$ and $e_2$ are related at the step index~$m$.
Note that, unlike when we chose a future world or extended lock state, we use \emph{strict} inequality here.
This guarantees that we only recursively check the relation for smaller step indices, ensuring wellfoundedness.

\subsection{The value relation}
\label{sec:value-relation}

\begin{figure*}
	\tiny
	\begin{mathpar}
		\begin{array}{ll}
			\vb{\AnnotateType{\AType}{\PolicyP}} &\triangleq \begin{array}[t]{l}
				\set{(v,v',W,\unimp{m}) \where \PolicyOrder{\PolicyP}{\AttackerA} \land (v,v',W,\unimp{m}) \in \vb{\AType}}\\
				\cup \set{(v,v',W,\unimp{m}) \where \NotPolicyOrder{\PolicyP}{\AttackerA} \land (v,W.\theta_1,\unimp{m}) \in \vs{\AType} \land (v',W.\theta_2,\unimp{m}) \in \vs{\AType}}
			\end{array}\\
			\vb{\Unit} &\triangleq \set{(\UnitVal,\UnitVal,W,\unimp{m})}\\
			\vb{\Nat} &\triangleq \set{(n,n,W,\unimp{m}) \where n \in \NN}\\
			\vb{\ProdType{\TypeTauI}{\TypeTauII}} &\triangleq \set{(\Pair{v_1}{v_2},\Pair{v'_1}{v'_2},W,\unimp{m}) \where (v_1,v'_1,W,\unimp{m}) \in \vb{\TypeTauI} \land (v_2,v'_2,W,\unimp{m}) \in \vb{\TypeTauII}} \\
			\vb{\SumType{\TypeTauI}{\TypeTauII}} &\triangleq \set{(\Inl(v), \Inl(v'), W,\unimp{m}) \where (v,v',W,\unimp{m}) \in \vb{\TypeTauI}} \\ &\cup \set{(\Inr(v), \Inr(v'), W,\unimp{m}) \where (v,v',W,\unimp{m}) \in \vb{\TypeTauII}} \\
			\vb{\RefType{\TypeTau}} &\triangleq \set{(l,l',W,\unimp{m}) \where W.\theta_1(l) = \TypeTau = W.\theta_2(l') \land (l,l') \in W.\beta} \\
			\vb{\ArrowType{\newstuff{\LockSetSigma}}{\pc}{\TypeTauI}{\TypeTauII}} &\triangleq \comprehend{(\Lam{\Var{x}}{e_1}, \Lam{\Var{x}}{e_2}, W,\unimp{m})}{
				\begin{array}{l}
					(\forall W'. W' \sqsupseteq W \rightarrow \unimp{\forall m'. m' < m} \rightarrow\\
					\forall v_1, v_2. (v_1,v_2,W',\unimp{m'}) \in \vb{\TypeTauI} \rightarrow \newstuff{\forall \LockSetSigmaI, \LockSetSigmaII. \LockSetSigmaI \supseteq \LockSetSigma \subseteq \LockSetSigmaII \rightarrow}\\
					(\subst{v_1}{e_1}, \subst{v_2}{e_2}, W', \newstuff{\LockSetSigmaI, \LockSetSigmaII}, \unimp{m'}) \in \eb[\pc]{\TypeTauII}) \ \land \\
					(\Lam{\Var{x}}{e_1},W.\theta_1,\unimp{m}) \in \vs{\ArrowType{\newstuff{\LockSetSigma}}{\pc}{\TypeTauI}{\TypeTauII}} \land\\
					(\Lam{\Var{x}}{e_2},W.\theta_2,\unimp{m}) \in \vs{\ArrowType{\newstuff{\LockSetSigma}}{\pc}{\TypeTauI}{\TypeTauII}}
			\end{array}	} \\
			\vb{\Gamma} &\triangleq \set{(\gamma,W,\unimp{m}) \where \Dom(\Gamma) \subseteq \Dom(\gamma) \land \forall \Var{x} \in \Dom(\Gamma). ( \gamma_1(\Var{x}),\gamma_2(\Var{x}),W,\unimp{m}) \in \vb{\Gamma(\Var{x})}}
		\end{array}
	\end{mathpar}
	\caption{Binary value relation}
	\label{fig:binary-value-rel}
\end{figure*}

We now consider the value relation, which can be found in Figure~\ref{fig:binary-value-rel}.
Here is where our relationship to the logical relation of FG~(\cite{RajaniG18}) stands out the most.
Unlike the expression relation, which needed significant changes to accommodate declassification, the value relation needs mostly cosmetic changes to account for declassification.
As before, we highlight these changes in \newtext{yellow}.
We also highlight, step indexes in \unimptext{gray} again. As before, a reader not familiar with step indexing may ignore them without missing any important insights. 

We start with a description of the first clause in Figure~\ref{fig:binary-value-rel}.
This clause defines the value relation for annotated types $\TypeTau$, which always have the form $\AnnotateType{\AType}{\PolicyP}$.
First, we ensure that the policy~$\PolicyP$ is respected, and then we use the type~$\AType$ to ensure indistinguishability.
To ensure that $\PolicyP$ is respected, we check whether $\PolicyOrder{\PolicyP}{\AttackerA}$.
If not, then $\AttackerA$ cannot see data labeled with~$\PolicyP$, so we simply check that the two expressions are independently in the unary relation (defined in Section~\ref{sec:unary-relation}), which
ensures that any functions nested inside the value do not contain implicit leaks.
This acts as a confinement lemma.
If $\PolicyOrder{\PolicyP}{\AttackerA}$ then $\AttackerA$ can see data labeled with~$\PolicyP$, so we check that the values are indistinguishable at the type~$\AType$.

For the base types~$\Unit$ and~$\Nat$, indistinguishability is equality on values of the type.
For product types~$\ProdType{\TypeTauI}{\TypeTauII}$, we check that both values are pairs.
Then, we compare the left and right projections of the values at their respective types.
For sum types~$\SumType{\TypeTauI}{\TypeTauII}$, we need to check that either both values are $\Inl$ or both values are $\Inr$ before checking at the respective summand type.

Two values of a reference type~$\RefType{\TypeTau}$ are related if they are locations related by $W.\beta$ and their types match those in~$W$.

Finally, two functions of type $\ArrowType{\LockSetSigma}{\pc}{\TypeTauI}{\TypeTauII}$ are indistinguishable if, when applied to two indistinguishable arguments, they produce indistinguishable results;
this definition is the secret sauce for a compositional definition of security.
We thus pick two related---and hence indistinguishable---values of type~$\TypeTauI$, called~$v_1$ and~$v_2$.
We then check that the bodies~$e_1$ and~$e_2$ of the functions, after substituting $v_1$ and $v_2$ for $x$, are related in the \emph{expression relation} at $\TypeTauII$.
This check requires that we know the open locks of both expressions.
While we do not know what the open locks will be when these functions are called, we know that they will be at least $\LockSetSigma$, the lock set given on the function type.
We similarly choose a future world and smaller step index, ensuring that our relation is monotonic.
Finally, in order to prevent functions that have side effects below $\pc$ from being related, we explicitly check that both functions are in the unary relation.

\subsection{Definition of security}
\label{sec:security-def}

Now that we have defined when an attacker~$\AttackerA$ cannot distinguish two programs up to relevant declassification, we use that definition to define when a program is secure.
Intuitively, a program is secure when no attacker can distinguish it from itself when given different secret inputs but the same public inputs, where secret and public respectively refer to values the chosen attacker cannot see and can see.
Since \calcname{} is a stateful functional language, secret inputs can be both the contents of secret locations of the initial state \emph{and} substitutions of free variables with secret types.
While we have discussed how we ensure that we consider programs with different secret parts of the state, we still need to substitute free variables.

To do so, we first define binary substitutions: a binary substitution~$\gamma \triangleq (\gamma_1,\gamma_2)$ is a pair of substitutions~$\gamma_1$ and~$\gamma_2$.
Given a context~$\Gamma$, we say that a binary~substitution~$\gamma$ is \emph{suitable for $\Gamma$} using the logical relation $\vb{\Gamma}$ defined in Figure~\ref{fig:binary-value-rel}.
Intuitively, a pair of substitutions~$(\gamma_1, \gamma_2)$ is suitable for $\Gamma$ if they substitute every variable $x : \TypeTau \in \Gamma$ with a pair of values~$(\gamma_1(x), \gamma_2(x))$ which are related at $\TypeTau$.
Since the value relation allows arbitrarily different values for secret types, this correctly allows differing secrets in different runs.
 
Given a program~$e$ and a suitable binary substitution~$\gamma$, the substituted programs~$\gamma_1(e)$ and~$\gamma_2(e)$ represent $e$ in two different runs with different secret inputs.
In order to deem $e$ secure, we ensure both that these two programs are always indistinguishable from themselves and that $\gamma_1(e)$ and $\gamma_2(e)$ are in the unary logical relation.
This guarantees that the type and $\pc$ are respected.
\begin{definition}[Security]\label{def:security}
  Let~$e$ be a program with free variables in~$\Gamma$ and locations in~$\theta$.
  Then we say that $e$ is \emph{secure at type~$\TypeTau$ with program counter~$\pc$ under open locks $\LockSetSigma$} if, for every attacker~$\AttackerA$, step index~$m$, world~$W \sqsupseteq (\theta, \theta, \text{id}_{\Dom(\theta)})$, and binary substitution~$\gamma$ such that \mbox{$(\gamma,W,m) \in \vb{\Gamma}$},
  $$\begin{array}{l}
      (\gamma_1(e), \gamma_2(e), W, \LockSetSigma, \LockSetSigma, m) \in \eb{\TypeTau}\ \land\\
      (\gamma_1(e), W.\theta_1, m) \in \es{\TypeTau}{\pc} \land (\gamma_2(e), W.\theta_2, m) \in \es{\TypeTau}{\pc}
    \end{array}$$
\end{definition}

To see why this corresponds to our intuitive notion of security, imagine that program~$e$ contains an illegal declassification.
In particular, imagine that some lock~$\LockSigma$ is required for the declassification, which happens without $\LockSigma$ open.
Then $e$ will not be related to itself for the attacker $(\A, \comprehend{\LockSigmaPrime}{\LockSigmaPrime \neq \LockSigma})$.

Also note that this definition finds information leaks that happen after a legitimate declassification. To see why, consider the following program:
$$ (\Open{\LockSigma}{\Assgn{l}{\Deref{h}}}); \Assgn{l'}{\Deref{h'}}$$
Assume that $h$ and $h'$ are protected by the lock $\LockSigma$ whereas $l$ and $l'$ are public.
One might be concerned that the legitimate declassification of $h$ to $l$ might mask the leak of $h'$ to $l'$ later in the program because we stop checking indistinguishability after relevant declassification.
However, our security property considers all pairs of states that are compatible with the initial world.
This includes, in particular, pairs of states $S$ and $S'$ in which the values of $h$ are the same but the values of $h'$ are different.
Assuming that $h$ and $h'$ point to integers, one might for example have $S(h) = 1$, $S(h') = 2$ and $S'(h) = 1$, $S'(h') = 0$.
For these $S$ and $S'$, the write of $h$ to $l$ would not constitute relevant declassification because the attacker cannot distinguish the two writes.
Hence, we will continue to enforce indistinguishability and find the leak of $h'$ via $l'$.
\footnote{This explanation contains a slight simplification.
  Technically we \emph{reset} the state after each step.
  This means we would \del{even }disallow a leak of $h$, even if $h$ was correctly declassified before. Consequently, our definition is not flow sensitive. We will briefly discuss this aspect in the conclusion.}

\mnb{\calcname{} types enforce security}
Now that we have formally defined when a program is secure, we would like to prove that our type system enforces program security.
This corresponds exactly to a standard theorem for logical relations, the \emph{fundamental theorem}.
The fundamental theorem says that well typed programs are self-related.
Since we defined security as self-relation, in our case the fundamental theorem tells us that all well typed programs are secure.
(The fundamental theorem for the unary logical relation, which implies only that the type system enforces the meaning of $\pc$ correctly, can be found in Section~\ref{sec:unary-relation}.)
\begin{theorem}[Binary fundamental theorem]
  \label{fundamental-binary}~\\
  If \mbox{$\Typing{\Gamma; \LockSetSigma; \theta}{\pc}{e}{\TypeTau}$}, $(\gamma,W,m) \in \vb{\Gamma}$, \mbox{$W.\theta_1 \sqsupseteq \theta$}, \mbox{$W.\theta_2 \sqsupseteq \theta$}, and for every location~$l~\in~\Dom(\theta)$, we have $(l,l) \in W.\beta$, then $\forall \AttackerA, m.~(\gamma_1(e),\gamma_2(e),W,\LockSetSigma,\LockSetSigma,m)~\in~\eb{\TypeTau}$.
\end{theorem}
We prove this theorem by induction on the given typing derivation of~$e$.
\added{This proof boils down to proving that each typing rule is semantically sound.
  In other words, we prove that each typing rule is still true even if we replace syntactic typing with self-relatedness in the binary relation (under a suitable binary substitution for free variables).
  In fact, in most cases it is simpler to prove a slightly stronger version that directly uses two different terms instead of obtaining them via a binary substitution.
  Let us consider the case for the \mbox{\rname{new}} rule as an illustrative example.
  Recall the typing rule for \NewKW:}
\begin{mathpar}
  \addedmaths{\Infer{new}
    {
      \Typing{\Gamma;\newstuff{\LockSetSigma};\theta}{\pc}{e}{\TauPrime}\\
      \PolicyOrder{\pc}{\TypeTau}\\
      \Subtyping{\newstuff{\PolicySpecialize{\TauPrime}{\LockSetSigma}}}{\TypeTau}
    }
    {\Typing{\Gamma;\newstuff{\LockSetSigma};\theta}{\pc}{\New(e,\TypeTau)}{\AnnotateType{(\RefType{\TypeTau})}{\PolicyBot}}}}
\end{mathpar}
\added{As a semantic analogue, we show the following compatibility lemma:}
\begin{lemma}
		\label{new-compatible}
		\added{
		If $(e,e',W,\LockSetSigma,\LockSetSigmaPrime,m) \in \eb{\TauPrime}$, $\LockSetSigma \lowequiv \LockSetSigmaPrime$ and $\Subtyping{\PolicySpecialize{\TauPrime}{\LockSetSigma}}{\TypeTau}$ and $\Subtyping{\PolicySpecialize{\TauPrime}{\LockSetSigmaPrime}}{\TypeTau}$, then $(\New(e,\tau),\New(e',\tau),W,\LockSetSigma,\LockSetSigmaPrime,m) \in \eb{\AnnotateType{(\RefType{\TypeTau})}{\PolicyBot}}$.}
\end{lemma}
\added{Note, that the $\pc$ is \emph{not} relevant here---it only plays a role in the unary relation---and so we do not need the assumption about the $\pc$ from the typing rule.
The proof of this compatibility lemma first considers the case where $(e,e',W,\LockSetSigma,\LockSetSigmaPrime,m)$ are in the reduction relation.
In this case, any reductions of $e$ or $e'$ correspond to a reduction of $\New(e,\tau)$ or $\New(e',\tau)$ via the \mbox{\rname{ENew}} reduction rule and the guarantees we get from the relatedness of $e$ and $e'$ via the $C_{\textsc{L}}$, $C_{\textsc{R}}$, or $C_{\textsc{Par}}$ clause are sufficient to prove relatedness via the corresponding clause for $\New(e,\tau)$ and $\New(e',\tau)$.

The more interesting case is if $e$ and $e'$ are values $v$ and $v'$.
In this case, any reduction of $\New(v,\tau)$ and $\New(v',\tau)$ happens via the \mbox{\rname{ENewBeta}} reduction rule producing write observation $\WriteObs{l}{\TauPrime}(v)$ and $\WriteObs{l}{\TauPrime}(v')$ for some new locations $l$ and $l'$.
We show that the two reductions are related via the $C_{\textsc{Par}}$ clause. 
The precondition of this clause is that there is no relevant declassification, which means we get two assumptions.
First, we can assume that the active lock sets of the reductions (say, $\LockSetSigmaPrimeI$ and $\LockSetSigmaPrimeII$) are below the capabilities of the attacker.
Second, we can assume that $\WriteObs{l}{\TauPrime}(v) \equivWorld[W',m'] \WriteObs{l}{\TauPrime}(v')$ for some world $W'$ extending $W$ and some $m' < m$.}
\added{
We then need to find a new world $W''$ extending $W'$, which we pick to be the world that extends the left and right state environments of $W'$ with the new locations $l$ and $l'$, respectively, and that adds the pair $(l,l')$ in the partial bijection.}

\added{We have to show that the resulting states after the reduction are related in this new world and that $\WriteObs{l}{\TauPrime}(v) \equivWelt[W',m'] \WriteObs{l}{\TauPrime}(v')$.
For both goals, the only interesting part is showing that $(l,l')$ are related in $W''.\beta$ and that $(v,v',W,\LockSetSigma,\LockSetSigmaPrime,m') \in \eb{\TypeTau}$.
Showing the locations are related is trivial due to the definition of $W''$.
But so far, we only know that $v$ and $v'$ are related at type $\TauPrime$, not at $\TypeTau$.
If $\TypeTau$ is secret, i.e., not visible to the attacker, we only need to show that $v, v'$ are in the unary relation individually.
This is straightforward, as the unary relation ignores the policy and we get what we need from the subtyping assumption.
If, however, $\TypeTau$ is visible to the attacker, we have to use the assumption that we do not have relevant declassification.
This either gives us $\WriteObs{l}{\TauPrime}(v) \equivWelt[W',m'] \WriteObs{l}{\TauPrime}(v')$ or the fact that either of the active lock sets are below the capabilities of the attacker.
In the first case, we get relatedness of $v$ and $v'$ from the derivation of $\WriteObs{l}{\TauPrime}(v) \equivWelt[W',m'] \WriteObs{l}{\TauPrime}(v')$.
In the second case, we know that the specialization in the subtyping is irrelevant, since it only removes locks the attacker can already open.
Hence, the specialization in the subtyping does not actually matter here, and we derive our goal from the soundness of subtyping.}

\added{
The case for $\NewKW$ is illustrative of the general form of proofs of the semantic soundness of most typing rules:
if a reduction step just reduces a subterm, we lift any guarantees for the subterm to the whole term.
For top-level reductions, we need to establish relatedness based on the specifics of the reduction step.
Usually, we have the same construct on both sides and we can establish relatedness via the $C_{\textsc{Par}}$ clause.
If the reduction produces \mbox{non-$\EmptyObs$}~observations, we need to reason about declassification, as in the case for $\NewKW$ discussed~above.}

\added{
An exception to this general proof setup occurs in rules which can branch (\mbox{\rname{case}} and \mbox{\rname{when}}).
When branching on a high value in those rules, we might need to establish that two entirely unrelated terms are indistinguishable.
In these cases, we rely on the following lemma, which establishes that any two secret terms are indistinguishable if they individually lie in the unary relation at a $\pc$ that is not visible to the attacker:
}
\begin{lemma}[\added{Equivalence of high expressions}]~\\
	\label{high-equiv}
	\added{If $(e,\theta,m)~\in~\es{\TypeTau}{\pc}$ and  $(e',\theta',m)~\in~\es{\TypeTau}{\pc}$ and $\NotPolicyOrder{\TypeTau}{\AttackerA}$ and $\NotPolicyOrder{\pc}{\AttackerA}$, then $ \forall \LockSetSigma,\LockSetSigmaPrime,\beta.\ \LockSetSigma~\lowequiv~\LockSetSigmaPrime \rightarrow (e,e',(\theta,\theta',\beta),\LockSetSigma,\LockSetSigmaPrime,m)~\in~\eb{\TypeTau}$.}
\end{lemma}
\added{
To prove this lemma, we exploit the guarantees given by the unary relation, which we discuss in more detail in Section \mbox{\ref{sec:unary-relation}}.
For now, it suffices to say that we can always use the $C_{\textsc{L}}$ clause for reductions of $e$ and the $C_{\textsc{R}}$ clause for reductions of $e'$ because the unary relation guarantees that any generated observation is above the $\pc$.
Because we assume the $\pc$ is at a level not visible to the attacker, by transitivity any produced observations are also not visible to the attacker.}

As a consequence of this proof \added{technique (proving semantic analogues of our syntactic typing rules)},  we can use the logical relation for \emph{semantic typing}, which
allows a programmer to compose a program which they have proven secure directly using the logical relation with a program proven secure using the type system, and obtain a proof that the composed program is secure.

% !TeX spellcheck = en_US
\subsection{The unary relation}
\label{sec:unary-relation}

\begin{figure*}
	\tiny
\begin{align*}
  &\es{\TypeTau}{\pc} &&\triangleq  \esb{\TypeTau}{\pc} \cup\ \vs{\TypeTau}\\
  &\esb{\TypeTau}{\pc} &&\triangleq \comprehend{(e,\theta,\unimp{m})}{
                          \begin{array}{l}
                            e \notin \mathcal{V} \land \forall S,\theta', \unimp{m'}, e', S', \ObservationOmega, \newstuff{\LockSetSigma,\LockSetSigmaPrime}.\ \theta' \sqsupseteq \theta \land \unimp{m' < m} \land (S,\unimp{m'}) \triangleright (\theta') \rightarrow \\
                            \Step{e,\newstuff{\LockSetSigma},S, e',S', \ObservationOmega, \newstuff{\LockSetSigmaPrime}}\ \rightarrow (\ObservationOmega = \EmptyObs \lor
                            \PolicyOrder{\pc}{\ObsPolicy(\ObservationOmega)})\ \land\\
                            (\exists \theta''.\ \theta'' \sqsupseteq \theta' \land (S',\unimp{m'}) \triangleright (\theta'') \land  (e',\theta'',\unimp{m'}) \in \es{\TypeTau}{\pc})
                          \end{array}}\\
 &\vs{\Unit} &&\triangleq \set{(e\UnitVal,\theta,\unimp{m})} \\
  &\vs{\Nat} &&\triangleq \set{(n,\theta,\unimp{m}) \where n \in \NN} \\
  &\vs{\ProdType{\TypeTauI}{\TypeTauII}} &&\triangleq \set{(\Pair{v_1}{v_2},\theta,\unimp{m}) \where (v_1,\theta,\unimp{m}) \in \vs{\TypeTauI} \land (v_2,\theta,\unimp{m}) \in \vs{\TypeTauII}} \\
  &\vs{\SumType{\TypeTauI}{\TypeTauII}} &&\triangleq \set{(\Inl(v),\theta,\unimp{m}) \where (v,\theta,\unimp{m}) \in \vs{\TypeTauI}} \cup \set{(\Inr(v),\theta,\unimp{m}) \where (v,\theta,\unimp{m}) \in \vs{\TypeTauII}} \\
  &\vs{\ArrowType{\newstuff{\LockSetSigma}}{\pc}{\TypeTauI}{\TypeTauII}} &&\triangleq \comprehend{(\Lam{\Var{x}}{e},\theta,\unimp{m})}{
                                                               \begin{array}{l}
                                                                 \forall \theta', v, \unimp{m'}. \theta' \sqsupseteq \theta \land \unimp{m' < m}\ \land \\ (v,\theta',\unimp{m'}) \in \vs{\TypeTauI} \rightarrow (\subst{v}{e}, \theta',\unimp{m'}) \in \es{\TypeTauII}{\pc}
                                                               \end{array}
                                                               } \\
  &\vs{\RefType{\TypeTau}} &&\triangleq \set{(l,\theta,\unimp{m}) \where \theta(l) = \TypeTau} \\
  &\vs{\AnnotateType{\AType}{\PolicyP}} &&\triangleq \vs{\AType}\\	
  &\vs{\Gamma} &&\triangleq \set{(\delta,\theta,\unimp{m}) \where \Dom(\Gamma) \subseteq \Dom(\delta) \land \forall \Var{x} \in \Dom(\Gamma). (\delta(\Var{x}),\theta,\unimp{m}) \in \vs{\Gamma(\Var{x})}}
\end{align*}
\caption{Unary relation}
\label{fig:unary-rel}
\end{figure*}

Our unary relation largely follows the work of \citet{RajaniG18}. Here, we discuss the relation briefly.
For readers familiar with FG, the differences from FG's unary logical relation are that:
(1) our relation is based on our small-step semantics while FG's relation uses a big-step semantics, and
(2) our relation adds locks sets at relevant places to match the revised syntax of types and the operational semantics, but this addition is only cosmetic since these sets are irrelevant for the unary relation.
Other than this, the unary relations are essentially the same.

The definition of the unary relation can be found in Figure~\ref{fig:unary-rel}.
Again, everything related to step-indexes is highlighted in \unimptext{gray}.
The additions for lock sets are marked in \newstuff{yellow}.

Much like the binary relation, the unary relation consists of an expression relation and a value relation.
Both relations are step- and world-indexed but the worlds are much simpler since we consider only one run at a time.
The simplified worlds are just state environments---$\theta$ in our notation.
A state environment~$\theta'$ is a future world of $\theta$, written $\theta' \sqsupseteq \theta$, if $\theta \subseteq \theta'$.

\begin{definition}[State well-formedness (\cite{RajaniG18})]
  We say that the state~$S$ is well-formed in the state environment $\theta$ at step index $m$, written $(S,m) \triangleright (\theta)$, if:
  \begin{itemize}
  \item[] $\Dom(\theta) \subseteq \Dom(S)$
  \item[$\land$] $(\forall l \in \Dom(\theta). (S(l),\theta,m) \in \vs{\theta(l)})$
  \item[$\land$] $\forall l \in \Dom(\theta). \theta(l) = \HeapTypeLookup(S,l)$
  \end{itemize} 
\end{definition}

The expression relation is again a union of a \added{reduction}\del{$\beta$} relation and the value relation.
The \added{reduction}\del{$\beta$} relation, and thus the expression relation as a whole, is indexed by a program-counter label~$\pc$.
The \added{reduction}\del{$\beta$} relation acts primarily as a confinement lemma, ensuring that any observation made during the execution of an expression has a policy above $\pc$, i.e., $\PolicyOrder{\pc}{\ObsPolicy(\ObservationOmega)}$.
A program $e$ is in the relation if for any starting state~$S$, $e$ can take a step to some $e'$ and a new state~$S'$, where the new state is correct for some future world $\theta'$ and the observation produced by the step, if non-trivial, has a policy above~$\pc$.

The value relation ignores the policy annotation on the type and ensures that values have the ``right'' shapes for their respective types.
Thus, values of type $\Unit$ and $\Nat$ are simply $\UnitVal$ and natural numbers, respectively,
sums and products recursively check constituting values against constituting types, and values of a reference type $\RefType{\TypeTau}$ are locations that have type $\TypeTau$ in the current world.
A function is in the relation of its type if it yields an expression of the result type when applied to any value of the input type.
Again, this makes the unary type relation compositional.

We prove the fundamental theorem for the unary logical relation as well.
\begin{theorem}[Unary Fundamental Theorem]~\\
  \label{fundamental-unary}
  If $\Typing{\Gamma; \LockSetSigma; \theta}{\pc}{\bar{e}}{\TypeTau}$ and $\theta' \sqsupseteq \theta$ and $(\delta,\theta',m) \in \vs{\Gamma}$, then $(\delta(\bar{e}
  ),\theta',m) \in \es{\TypeTau}{\pc}$.
\end{theorem}

\added{As a consequence of the two fundamental theorems we get that our type system enforces security:}
\begin{theorem}[Security]
	\added{
	If \mbox{$\Typing{\Gamma; \LockSetSigma; \theta}{\pc}{e}{\TypeTau}$}, then $e$ is secure at type~$\TypeTau$ with program counter~$\pc$ under open locks $\LockSetSigma$.}
\end{theorem}

% !TeX spellcheck = en_US
\section{Comparison to Flow-Lock security}
\label{sec:security}

We use the declassification mechanism of Flow Locks~(\cite{BrobergS09}) and our security definition is inspired by Flow-Lock security.
It is therefore reasonable to ask if the two definitions are also semantically related and, if so, whether this relationship can be made mathematically~precise.

Upfront, the two definitions are \emph{not equivalent}.
Our definition is motivated by the desire for a compositional security property, which means that we consider some functions to  be insecure.
Since Flow Locks considers all functions secure, our security definition is more restrictive.
However, our security definition \emph{implies} Flow-Lock security.

In order to compare our guarantees to those of Flow Locks, we have to translate the security definitions of Flow Locks to \calcname{}.
Because our language is more complex than that of Flow Locks, this translation reasonably extends the definitions of Flow Locks to cover our language constructs.
We then use our logical relation to show that secure \calcname{} programs are also Flow-Lock secure.
However, even the extended Flow-Lock security definition does not apply to programs with higher-order state, so we must restrict the formal theorem relating the two definitions (Theorem~\ref{thm:flow-locks-security}) to \calcname{} programs without higher-order state.

\paragraph{Flow Locks defines security using \textbf{attacker knowledge}}
Here is how Flow-Lock security works intuitively.
Consider two executions that start in attacker-indistinguishable states and have attacker-indistinguishable observations up to some point in the execution.
Then, at the next step, either there is a relevant declassification or the attacker \emph{learns nothing new about the initial state}, i.e., the observations remain attacker-indistinguishable.
The rest of this section formalizes this intuitive understanding of Flow-Lock security and explains the relation between that and our security as defined in Section~\ref{sec:comp-defin-secur}.

We already have a way to determine when two states are indistinguishable to an attacker~$\AttackerA$ up to relevant declassification, given via the logical relation in Definition~\ref{def:state-indist}.
However, in order to match the development of Flow Locks, we need to be more precise.

\added{To be able to talk about the parts of a state $\AttackerA$ can see, we follow \mbox{\cite{BrobergS09}} and define the $\AttackerA$-low version of a state $S$:}
$$\addedmaths{\low{S} \triangleq \{\HeapMapsTo{l}{v,\tau} \in S \mid \PolicyOrder{\TypeTau}{\AttackerA}\}}$$
\added{
We say a state $L$ is $\AttackerA$-low if it contains only $\AttackerA$-visible locations, that is if $\low{L} = L$.
As a notational convention, we use the captial letter $L$ for $\AttackerA$-low states.}

\added{	
We also define a state environment $\theta_S$ corresponding to a given state: $\theta_S(l)~:=~\HeapTypeLookup(S,l)$}
\added{Now we can define low equivalence of two states:}
\del{We consider states~$S_1$ and $S_2$ with the same set of low locations (i.e., locations which $\AttackerA$ can see).
More precisely, we assume that we have some state environment~$\theta$ which agrees with $S_1$ and $S_2$ on the types of those low locations, so $\Dom(\theta)$ is the set of low locations in $S_1$ and $S_2$.}
\del{Then, w}\added{W}e say that $S_1$ and $S_2$ are \emph{low~equivalent}, written $S \lowequiv S'$,
\added{if they have the same $\AttackerA$-low locations (i.e., locations which $\AttackerA$ can see) and} if \added{they are related at every step index}\del{for every step index $m$, }$\delmaths{(S_1, S_2, m) \bwf (\theta, \theta, \text{id}_{\Dom(\theta)})}$.
\begin{mathpar}
	 \addedmaths{S_1 \lowequiv S_2\triangleq
	 	\begin{array}{l}
	 	\Dom(\low{S_1}) = \Dom(\low{S_2})~\land\\
	 	 \forall m.\ (\low{S_1},\low{S_2},m) \bwf (\theta_{\low{S_1}},\theta_{\low{S_2}},id_{\Dom(\low{S_1})})
	 	\end{array}
 	 }
	
\end{mathpar}
Note that this definition imposes no requirements on the high locations.

We refer to lists of observations which result from consecutive steps as \emph{traces}.
For example, if~$e$ steps to~$e'$ generating observation~$\ObservationOmega$ and $e'$ steps to $e''$ generating observation~$\ObservationOmega'$, then the resulting trace is $\ObservationOmega,\ObservationOmega'$.
We further write \mbox{$\Steps{e,S,\LockSetSigma,\PTraceOmega,\AttackerA,e',S',\LockSetSigmaPrime}$} when $e$ reduces to $e'$ in state~$S$ and lock state~$\LockSetSigma$ over several steps, resulting in new state~$S'$ with active lock set~$\LockSetSigmaPrime$ in the final step, allowing attacker~$\AttackerA$ to make observations~$\PTraceOmega$.
\added{Formally, $\AttackerA$-observable traces are defined as follows.}

\begin{definition}[\added{$\mathcal{A}$-visible trace reduction}]
	\begin{mathpar}
		\addedmaths{
		\nored{e}{S}{\LockSetSigma}} \and
		\addedmaths{
		\Infer{inv-red}{\Steps{e,S,\LockSetSigma,\PTraceOmega,\AttackerA, e'',S'',\LockSetSigmaPrime} \\ \Step{e'',\LockSetSigma,S'', e',S',\ObservationOmega,\LockSetSigmaDoublePrime} \\ \inv{\ObservationOmega}}{\Steps{e,S,\LockSetSigma,\PTraceOmega,\AttackerA, e',S',\LockSetSigmaPrime}}}
		 \and
		 \addedmaths{
		\Infer{vis-red}{\Steps{e,S,\LockSetSigma,\PTraceOmega,\AttackerA,e'',S'',\LockSetSigmaDoublePrime} \\ \Step{e'',\LockSetSigma,S'', e',S',\ObservationOmega,\LockSetSigmaPrime} \\ \neg \inv{\ObservationOmega}}{\Steps{e,S,\LockSetSigma, {\PTraceOmega,\ObservationOmega}, \AttackerA, e',S',\LockSetSigmaPrime}}
	}
	\end{mathpar}
	\added{where $\inv{\ObservationOmega} \triangleq \NotPolicyOrder{\ObsPolicy(\ObservationOmega)}{\AttackerA} \lor \ObservationOmega = \EmptyObs$.}
\end{definition}

Note that such traces only contain attacker visible observations, so in particularly they do not explicitly contain $\EmptyObs$-observations.
We lift observation indistinguishability (Figure~\ref{fig:obs-indist}) to trace indistinguishability pointwise.

Finally, we can define the \emph{knowledge} of an attacker~(\cite{AskarovS07B}).
Intuitively, if an attacker observes the execution of some expression~$e$, then the knowledge of the attacker is the set of possible states that the execution could have started in.
Assume that the starting state is low equivalent to some state~$L$ which contains only low locations.
\del{Further assume that a state environment~$\theta_L$ agrees with~$L$ on the types of every location.}
We can then define attacker~$\AttackerA$'s~knowledge as follows:
{ \small
  $$
  \Know{\AttackerA}(e,\PTraceOmega,L,\LockSetSigma) \triangleq
  \left\{S \where \begin{array}{l}
                    S \lowequiv L\\
                    {}\mathrel{\land} \Steps{e,S,\LockSetSigma,\PTraceOmegaPrime,\AttackerA,e',S',\LockSetSigmaPrime}\\
                    {}\mathrel{\land} \ \exists W \supseteq (\theta_{L},\addedmaths{\theta_{S}},\text{id}_{\Dom(L)})\\
                    \hspace{1em}\mathrel{\textrm{such that}} \forall m.\ \PTraceOmega \equivWelt[(W,m)] \PTraceOmegaPrime
             \end{array}\!\right\}
  $$
}

Note that we use the logical relations based indistinguishability on observations here.
This definition coincides with the definition of indistinguishability used in Flow locks (\cite{BrobergS09}) for those observations present in \del{\mbox{\cite{BrobergS09}}}\added{Flow locks}.
Additionally, our definition extends to other types in our language in a way that is consistent with our treatment~of~indistinguishability.

The knowledge definition above is intuitive, but it behaves strangely if there is higher-order state, which allows storing nonterminating functions in memory.
To see this, \added{it is helpful to know how knowledge is used in the actual definition of flow lock security.
As explained earlier, the definition compares two executions of the same program, which \mbox{\cite{BrobergS09}} call \emph{Runs}.}
\added{
	Defining runs requires us to extend a state environment with additional types from a state.
	For a state environment $\theta$ and state $S$ we define a state environment $\theta^S : dom(\theta) \cup dom(S) \rightarrow Type$.}
\begin{align*}
	\addedmaths{\theta^S(l) :=} \begin{cases}
		\addedmaths{\theta (l)} & \addedmaths{\text{if $l \in \Dom(\theta)$}} \\
		\addedmaths{\HeapTypeLookup(S,l)} & \addedmaths{\text{otherwise}}
	\end{cases}
\end{align*}
%\added{An $\mathcal{A}$-observable run is defined in the following way:}
\[
\addedmaths{
	\mathit{Run}_\AttackerA(e,L,\LockSetSigma,\theta) := \set{(\PTraceOmega;\LockSetSigmaPrime) \where \exists S,S',e'.\ S \wf[\cdot] \theta^L \land S \lowequiv L \land \Steps{e,S,\LockSetSigma, \PTraceOmega, \AttackerA, e',S',\LockSetSigmaPrime}}}
\]
\added{where $S \wf[\cdot] \theta^L$ is a syntactic (and therefore step-index free) version of the unary state well formedness predicate $(S,m) \triangleright (\theta^L)$ replacing the invocation of the unary relation with a check that the program is well typed in the type system.
}

\added{Equipped with this notion, we can define a version of termination-insensitive Flow lock security (\mbox{\cite{BrobergS09})} for \calcname.}
\begin{definition}
	\added{A program $e$ is termination-insensitive flow lock secure if for all pairs of runs $(\PTraceOmega,\ObservationOmega;\LockSetSigmaPrime) \in \mathit{Run}_\AttackerA(e,L,\LockSetSigma,\theta)$ and $(\PTraceOmegaPrime,\ObservationOmegaPrime;\LockSetSigmaDoublePrime) \in \mathit{Run}_\AttackerA(e,L,\LockSetSigma,\theta)$ such that the initial parts of the traces are indistinguishable to $\AttackerA$---i.e.,~if $\forall m.\ \PTraceOmega \equivWorld[((\theta,\theta,\text{id}_{\Dom(\theta)}),m)] \PTraceOmegaPrime$---and either $\LockSetSigmaPrime \sqsubseteq \mathcal{A}$ or $\LockSetSigmaDoublePrime \sqsubseteq \mathcal{A}$, then $\Know{\AttackerA}(e,{\PTraceOmega,\ObservationOmega},L,\LockSetSigma) = \Know{\AttackerA}(e,{\PTraceOmegaPrime,\ObservationOmegaPrime},L,\LockSetSigma)$.}
	\end{definition}
\added{We now return to the question of knowledge-based security in the higher-order setting.}
Consider the program $\Assgn{l}{\Deref{l}}$ where $l$ is a low location.
Since this assigns a low value to a low location, it should intuitively be safe.
Moreover, our logical relation correctly deems this program secure.
However, knowledge-based security with the knowledge definition above would consider this program insecure.

This discrepancy comes from an assumption implicit in the definition of knowledge-based security: that the indistinguishability relation used on observations is an equivalence relation.
However, because our logical relation represents termination-\emph{in}sensitive indistinguishability, diverging functions are related to many functions that are not necessarily related to each other.
For example, every silently diverging function is indistinguishable from every other function.
\added{Now consider an $\AttackerA$-low state $L$}
\del{So the knowledge (set) of $\Assgn{l}{\Deref{l}}$ contains $\Lam{x}{\Const{5}}$ when starting in a state} that maps $l$ to a silently nonterminating function $f$\del{,}\added{.}
\added{Then $\mathit{Run}_\AttackerA(\Assgn{l}{\Deref{l}},L,\EmptyLock,\theta)$ contains a run writing $\Lam{x}{\Const{6}}$ to $l$ and one writing $f$ to $l$.
Since a write of $f$ is indistinguishable from a write of $\Lam{x}{\Const{5}}$, the knowledge for the run writing $f$ contains a state in which $l$ contains $\Lam{x}{\Const{5}}$.
On the other hand, the knowledge for the run writing $\Lam{x}{\Const{6}}$ to $l$ does not contain a state in which $l$ contains $\Lam{x}{\Const{5}}$ because $\Lam{x}{\Const{6}}$ and $\Lam{x}{\Const{5}}$ are distinguishable.
So the knowledge-based definition rejects this program.}
\del{
but not when starting in a state that maps $l$ to $\Lam{x}{\Const{6}}$, even though $f$ and $\Lam{x}{\Const{6}}$ are indistinguishable.}

If we had defined a termination-sensitive logical relation and additionally proved it an equivalence relation\footnote{Proving a step-indexed logical relation is an equivalence relation is generally very difficult even if the relation is termination sensitive.
Methods to do this in the general case are not known yet.}, we would not have run into this limitation of knowledge-based definitions.
However, we choose to stay with a termination-insensitive logical relation because:
(a) It is standard in literature.
In particular, the two security definitions of Flow Locks in \cite{BrobergS09} are termination-insensitive, and
(b) Any type system that is sound with respect to a termination-sensitive logical relation must also establish program termination, which is difficult in practice.

Hence, to get a meaningful comparison between Flow-Lock security and our security definition, we only consider programs with lower-order state, i.e., programs that do not store functions in memory.
This eliminates the problem outlined above.
This setting is still more general than that of Flow-Lock security, which does not consider any higher order programs, excluding even those higher-order programs that never store functions in memory.
With this restriction in place, we can prove that our lifting of Flow-Lock security (\cite{BrobergS10,BrobergS09}) is implied by our notion of security (for \calcname{}).

\begin{theorem}[Our security implies Flow-Lock security\del{[ sketch only; see the technical appendix for details{]}}]\label{thm:flow-locks-security}
  Let~$e$ be a program that does not use higher-order state, $\theta$ an arbitrary state environment, \added{and }$L$ a\added{n} \addedmaths{\AttackerA}-low state\del{, and $\theta^L$ the extension of $\theta$ so that it agrees with $L$ on the types of all locations in $L$}.
  \del{Imagine that}\added{If} $(e,e,(\theta^L,\theta^L,\text{id}_{\Dom(\theta^L)}),\LockSetSigma,\LockSetSigma,m) \in \eb{\TypeTau}$ at all step indexes $m$ and
  \added{$(\PTraceOmega,\ObservationOmega;\LockSetSigmaPrime)~\in~\mathit{Run}_\AttackerA(e,L,\LockSetSigma,\theta)$ and $(\PTraceOmegaPrime,\ObservationOmegaPrime;\LockSetSigmaDoublePrime) \in \mathit{Run}_\AttackerA(e,L,\LockSetSigma,\theta)$}
  \del{that an attacker~$\AttackerA$ executes~$e$ in two starting states that~$\AttackerA$ cannot distinguish from~$L$.
  This results in two $\AttackerA$-visible traces, $\PTraceOmega,\ObservationOmega$ and $\PTraceOmegaPrime,\ObservationOmegaPrime$.
  Let $\LockSetSigmaPrime$ and $\LockSetSigmaDoublePrime$ be the active lock sets in the steps producing $\ObservationOmega$ and $\ObservationOmegaPrime$, respectively.}
  \del{If}\added{and i}f the two initial traces are indistinguishable to $\AttackerA$---i.e.,\ \del{if }$\forall m.\ \PTraceOmega \equivWorld[((\theta,\theta,\text{id}_{\Dom(\theta)}),m)] \PTraceOmegaPrime$---and either $\LockSetSigmaPrime \sqsubseteq \mathcal{A}$ or $\LockSetSigmaDoublePrime \sqsubseteq \mathcal{A}$, then $\Know{\AttackerA}(e,{\PTraceOmega,\ObservationOmega},L,\LockSetSigma) = \Know{\AttackerA}(e,{\PTraceOmegaPrime,\ObservationOmegaPrime},L,\LockSetSigma)$. 
\end{theorem}

% !TeX spellcheck = en_US
\section{Related work}
\label{sec:related-work}

\mnb{Flow~Locks and Paralocks}
Flow~Locks and Paralocks form the basis of the declassification mechanism we present here. They also inspired our security definition.
The first work on Flow Locks~(\cite{BrobergS06}) introduces an unscoped version of the \OpenKW{} and \CloseKW{} constructs.
There are two very different security definitions for Flow Locks: a bisimulation-based definition~(\cite{BrobergS06}) and a knowledge-based definition~(\cite{BrobergS09}) (which we build on).
Our security definition generalizes Flow Locks' security definition to make it compositional, and to lift it to the higher-order setting with some technical changes, e.g., our \OpenKW{} and \CloseKW{}~constructs~are~scoped.

Paralocks~(\cite{BrobergS10}) is an extension of Flow Locks which both generalizes Flow-Locks policies
in several ways
and adds a \WhenKW construct that allows programs to test whether a lock is open.
Defining security for a language with \WhenKW requires making locks observable.
In order to focus on the key features of higher-order where declassification, we did not consider \added{most of }Paralocks' additions.
\del{However, the language of the accompanying technical appendix covers the \WhenKW~construct.}

\mnb{Higher-order where declassification}
\label{p:higher-where-decl}
We are only aware of two other works on where declassification for higher-order languages: the original Flow-Locks paper~(\cite{BrobergS06}) and \citeauthor{MatosB05}'s work on the non-disclosure policy~(\cite{MatosB05}).
Both papers give a bisimulation-based definition in the store-based setting which only guarantees the lack of information leaks during the computation of an expression to a value, without giving any guarantees about later uses of the resulting value.
This makes the security definitions noncompositional. 

\citet{BrobergS06} do not give security guarantees to their surface language.
Instead, they translate all programs to a restricted language where all programs are in A-normal form, and then give security guarantees for \emph{that} language.
They state that their original language ``[is] not well suited when defining the semantic security property,'' suggesting that their security property does~not~generalize.

\citet{MatosB05} use scoped flow constructs, similar to \calcname{}'s \OpenKW{} and \CloseKW{}.
This seems to be one of the key reasons that non-disclosure is usually defined as a where property, while Flow-Lock and Paralock~security are usually defined as \emph{when} properties.
However, they treat functions imprecisely in at least two ways.
First, they require that the output of a function be at least as sensitive as all of its inputs, even those on which the output does not depend, creating label creep.
Second, they define function indistinguishability syntactically, which leads to noncompositionality in their definition.

\mnb{Logical relations and information flow}
We build directly on the language, associated type system, and logical relation for~FG~(\cite{RajaniG18}).
FG does not allow any declassification.
Our work can be viewed as an extension of FG with where declassification, with policies specified in the style of Flow Locks.

\citet{GregersenBTB21} use logical relations to provide a noninterference guarantee to a language with higher-order state and polymorphism.
\cite{FruminKB21} go in a different direction with SeLoC by including concurrency.
Both papers use Iris~(\cite{JungSSSTBD15,JungKJBBD18}), a Coq~(\cite{Coq22}) framework for reasoning about software using separation logic, for their proofs and definitions.

While we believe that we are the first to use logical relations to reason about \emph{where} declassification,
logical relations have been used to reason about \emph{what} declassification.
\citet{CruzRST17} % hack to make things look right.
propose type-based relaxed noninterference, an object-oriented version of relaxed noninterference~(\cite{PengZ05}).
They use a logical relation to make type-based relaxed noninterference compositional, thus allowing relaxed noninterference to be lifted to object-oriented programming, which is inherently higher-order.
Likewise, \citet{NgoNR20} use a logical relation to adapt relaxed noninterference to the simply typed lambda calculus.

\added{\mbox{Recently, \cite{RajaniCK25}} built a modal type system and logical-relations--based semantic model to reason about \emph{what} declassification in the coarse-grained setting (where only some terms are explicitly annotated with a security label). 
In their model, they employ two graded monads, one to annotate a type with a security label, and one to allow declassification.
As in relaxed noninterference, declassification policies are functions that provide access to secrets.
\mbox{\cite{RajaniCK25}} distinguish between two types of declassification policies and treat them differently:
The first type allows use of a secret via a noninterfering declassification function.
The second type allows declassification via a declassification function that actually reveals information.
The first type of policy needs to be self-related in the binary relation, witnessing that it does not leak information, while the second kind of policy only needs to be in the unary relation.}

\mnb{Relevant declassification}
While we are the first to use relevant declassification as the basis of a model of types---and the first to name it---the insight that in the presence of (intensional) declassification we cannot always require indistinguishability is not new.
Our concrete definition of relevant declassification is inspired by the definitions of Flow-Lock and Paralock~security.
Specifically, the proofs in the appendixes and full developments of the respective papers~(\cite{BrobergS09,BrobergS10}) contain ideas that influenced various aspects of relevant declassification.

The closest analogues to our relevant declassification condition can be found in bisimulation-based definitions for declassification, which only require that the second execution match the first if some precondition is satisfied.
Often, this precondition requires that the memories the two programs are executed in are indistinguishable given the currently active declassification policies (in our case, the open locks).
This is, for example, the case in the nondisclosure policy~(\cite{MatosB05}) and the original Flow-Locks security definition~(\cite{BrobergS06}).
Note that, unlike our definition, this condition does \emph{not} require any difference in the observed behavior.

Our formalization of relevant declassification is also similar to conditions in the definition of strong $D$-bisimulations by \citet{MantelS04}.
\citeauthor{MantelS04} instrumented their operational semantics to distinguish steps in which declassification is allowed from other execution steps.
They then define a bisimulation which, similar to our logical relation, requires that declassifications happen in lockstep. 
Then, if a declassification step satisfies a condition very similar to our relevance condition, they do not require the resulting memories to be related. 
Concretely, their condition checks that (a) the downgrade is allowed by the declassification policy, (b) that the downgrade is visible at the currently considered security level (i.e, to the attacker), and (c) that the starting memories differ when viewed from the security level the declassification starts at.
Notably, in case of relevant declassification\added{, while they allow the resulting \emph{memories} to be different,} they still require the resulting \del{programs}\added{\emph{programs}} to be indistinguishable from each other.
We cannot do this without breaking compositionality.

Unlike our security condition, strong D-bisimulations are a store-based approach to security that cannot handle dynamic allocation of memory locations. They were applied to a first-order imperative language only.

The security definitions for escape-hatch style what-declassification mechanisms à la delimited release (\cite{SabelfeldM03}) contain a clause that might look superficially similar to our relevant declassification mechanism.
The clause ensures that indistinguishability is only enforced if the declassified expressions are indistinguishable in the two executions under consideration.
This condition is, however, fundamentally different from our notion of relevant declassification in that it can be checked \emph{once and for all} right at the beginning of the program.
This works because what declassification is an extensional property.
We have to make the relevance check at all writes during the program's executions because, in our setting, the interpretation of the policy and, hence, relevance, depends on the part of the program that is executing.

\mnb{Small-step logical relations}
In order to be able to give security guarantees for partial traces and diverging programs, we chose to give security guarantees for individual steps of computation, requiring that the resulting programs be related in the expression relation again.
While this \emph{small-step} approach was inspired by bisimulations, it is also standard in logical-relations models.
Notably, most logical relations built in Iris~(\cite{JungSSSTBD15,JungKJBBD18}) are defined using a weakest-precondition predicate defined over small-step reductions.
Most closely related to our definition are the logical relations of Simuliris~(\cite{GaeherSSJDKKD22})---which sets up a one-directional simulation between executions---and the previously mentioned SeLoC~(\cite{FruminKB21})---which sets up a strong lockstep bisimulation for noninterference, similar to our $C_{\mathsf{PAR}}$ clause (but without declassification).

\mnb{Knowledge-based security definitions}
\citet{AskarovS07B} first proposed knowledge-based security guarantees for \emph{gradual release}, which allows declassification via specific commands.
First, they give a knowledge-based definition of noninterference: attacker knowledge does not increase during program execution.
They then contrast it with a guarantee for gradual release itself: attacker knowledge may increase only at specific declassification commands.

Since then, knowledge-based definitions have been used by many researchers.
Importantly, \citet{BrobergS10} replaced their original bisimulation-based security definition for Flow Locks~(\cite{BrobergS06}) with a \emph{flow-sensitive} knowledge-based definition.
While our security definition is not flow sensitive, we believe that higher-order flow-sensitive security definitions should be possible through more advanced logical~relations~technology~(\cite{HurDNV12,AhmedDR09}).

All knowledge-based security definitions we are aware of apply to lower-order languages only.
As we noted in Section~\ref{sec:security}, knowledge-based security definitions assume that indistinguishability of observations is an equivalence relation.
It is unclear whether one can satisfy this assumption to get a termination-insensitive security definition in the presence of higher-order state.

%\input{future-work}
% !TeX spellcheck = en_US
\section{Conclusion}
\label{sec:conclusion}

We have defined \calcname{}, a higher-order language based on FG~(\cite{RajaniG18}) with where-declassification policies based on Flow Locks~(\cite{BrobergS09,BrobergS06}).
We were able to give a \emph{compositional} definition of security for \calcname, the first such definition for higher-order where declassification.
To achieve this, we used a logical relation, which deems two functions indistinguishable if indistinguishable inputs lead to indistinguishable results.
This definition is compositional on the language syntax.
Relevant declassification enabled us to reason about the intensional aspects of where declassification in our logical relation.
Formalizing and incorporating relevant declassification into the logical relation forms the core technical contribution of our work.

Although we are not the first to notice that logical relations are particularly well-suited to defining compositional security properties---they have been used both for noninterference~(\cite{RajaniG18,GregersenBTB21,FruminKB21}) and what declassification~(\cite{CruzRST17,NgoNR20}),
we have demonstrated a new use of logical relations by showing that they can be used to reason about intensional security~properties.

Our security definition is based on the lower-order definition of~\cite{BrobergS09}.
We lift their definition to the higher-order setting by embedding the key conditions in the expression relation of our logical relation.
We believe that this approach can be applied to other declassification mechanisms and styles as well.
For example, we believe that lifting bisimulation-based security definitions for other declassification mechanisms like those of \citet{MatosB05} and \citet{MantelS04} to the higher-order setting should pose few further technical challenges.
As noted in Section \ref{sec:related-work}, many bisimulation-based security definitions rely on conditions similar \added{to }our condition defining relevant declassification.
Our expression relation essentially \emph{is} a bisimulation that puts additional requirements on values via the value logical relation.
We therefore expect that we could similarly extend other bisimulation-based definitions to the higher-order setting by extending them with a value relation and replacing their notion of indistinguishability with relatedness in the logical relation, thus obtaining compositionality.

Furthermore, a logical-relations definition of security for \emph{when} declassification based on our relation should be straightforward.

Even though our definition of security is not flow-sensitive, this is not inherent in our approach.
By tracking the \emph{contents} of the state in the world, as is routine in Kripke logical relations~(\cite{AhmedDR09}), we would obtain a flow-sensitive security definition.
\citet{FruminKB21} and \citet{GregersenBTB21} already incorporate this in their logical relations for noninterference.

\section{Statement of competing interests}
The authors were funded by their respective academic employers during the execution of this work. There are no related external funding sources, financial interests or patents.

%% Bibliography
\bibliographystyle{ACM-Reference-Format}
\bibliography{sources}

\end{document}